\documentclass[12pt]{iopart}
\usepackage{graphicx}
\usepackage{xcolor}
\usepackage{mathrsfs}
\usepackage{subcaption}
\usepackage{enumitem}
\usepackage{bm}

\usepackage{iopams}  
\begin{document}

\title[Photon Calibration Performance of KAGRA during the 4th Joint Observing Run (O4)]{Photon Calibration Performance of KAGRA during the 4th Joint Observing Run (O4)}

\author{Dan Chen$^1$, Shingo Hido$^2$, Darkhan Tuyenbayev$^2$, Dripta Bhattacharjee$^3,^4$, Nobuyuki Kanda$^5$, Richard Savage$^6$, Rishabh Bajpai$^1$, Sadakazu Haino$^7$, Takahiro Sawada$^2$, Takahiro Yamamoto$^2$, Takayuki Tomaru$^1$, Yoshiki Moriwaki$^8$}

\address{$^1$ Gravitational Wave Science Project, National Astronomical Observatory of Japan, 2-21-1 Osawa, Mitaka, Tokyo 181-8588, Japan}
\address{$^2$ Institute for Cosmic Ray Research (ICRR), The University of Tokyo, 238
Higashi-Mozumi, Kamioka-cho, Hida City, Gifu 506-1205, Japan}
\address{$^3$ Kenyon College, Gambier, USA}
\address{$^4$ Missouri University of Science and Technology, Rolla, USA}
\address{$^5$ Nambu Yoichiro Institute of Theoretical and Experimental Physics (NITEP), Osaka Metropolitan University, 3-3-138 Sugimoto-cho, Sumiyoshi-ku, Osaka City, Osaka 558-8585, Japan}
\address{$^6$ LIGO Hanford Observatory, Richland, Washington, USA}
\address{$^7$ Institute of Physics, Academia Sinica, 128 Sec. 2, Academia Rd., Nankang, Taipei 11529, Taiwan}
\address{$^8$ Faculty of Science, University of Toyama, 3190 Gofuku, Toyama City, Toyama 930-8555, Japan}
\ead{dan.chen@nao.ac.jp, shingo@icrr.u-tokyo.ac.jp}
\vspace{10pt}
\begin{indented}
\item[]April 2025
\end{indented}

\begin{abstract}
\par KAGRA is a kilometer-scale cryogenic gravitational-wave (GW) detector in Japan.
It joined the 4th joint observing run (O4) in May 2023 in collaboration with the Laser Interferometer GW Observatory (LIGO) in the USA, and Virgo in Italy.
After one month of observations, KAGRA entered a break period to enhance its sensitivity to GWs, and it is planned to rejoin O4 before its scheduled end in October 2025.
\par To accurately recover the information encoded in the GW signals, it is essential to properly calibrate the observed signals.
We employ a photon calibration (Pcal) system as a reference signal injector to calibrate the output signals obtained from the telescope. In ideal future conditions, the uncertainty in Pcal could dominate the uncertainty in the observed data.
In this paper, we present the methods used to estimate the uncertainty in the Pcal systems employed during KAGRA O4 and report an estimated system uncertainty of 0.79\ \%, which is three times lower than the uncertainty achieved in the previous 3rd joint observing run (O3) in 2020.
Additionally, we investigate the uncertainty in the Pcal laser power sensors, which had the highest impact on the Pcal uncertainty, and estimate the beam positions on the KAGRA main mirror, which had the second highest impact.
\par The Pcal systems in KAGRA are the first fully functional calibration systems for a cryogenic GW telescope.
To avoid interference with the KAGRA cryogenic systems, the Pcal systems incorporate unique features regarding their placement and the use of telephoto cameras, which can capture images of the mirror surface at almost normal incidence.
As future GW telescopes, such as the Einstein Telescope, are expected to adopt cryogenic techniques, the performance of the KAGRA Pcal systems can serve as a valuable reference.

\end{abstract}

%
\vspace{2pc}
\noindent{\it Keywords}: gravitational wave, KAGRA, calibration, photon calibration system
%
%
%
%

\section{Introduction}
\par Gravitational-wave (GW) observation began with the binary black hole merge event that was observed by the Laser Interferometer GW Observatory (LIGO) in 2015 \cite{PhysRevLett.116.061102}.
Since then, LIGO, Virgo, and KAGRA, which are large-scale interferometer GW observatories located in the USA, Italy, and Japan, respectively, have comprised a GW observation network known as the LIGO–Virgo–KAGRA (LVK) collaboration.
This network detected 90 GW events in total by the end of O3 \cite{PhysRevX.13.041039} and is expected to detect many GW events during the 4th joint observing run (O4).
There was a mid-run break at the beginning of 2024 for three months, resulting in two O4 periods (O4a and O4b).
KAGRA joined O4a since its beginning for four weeks, and it is planned to join O4b at the beginning of 2025.
\par As the sensitivity of detectors increases, the importance of the observation data calibration also increases.
The photon calibration (Pcal) system is used as the main reference signal generator for data calibration in large-scale interferometer detectors.
LIGO, Virgo, and KAGRA employ their own Pcal systems \cite{10.1063/1.4967303}, \cite{Estevez_2021}, \cite{10.1063/5.0147888}.
\par As the Pcal system provides the reference signals to the interferometer, its uncertainty must be carefully examined because ultimately it could affect the parameter estimation of GW event sources.
This study demonstrates the Pcal uncertainty in KAGRA. Additionally, new estimation methods that reduce the Pcal uncertainty in other/future  GW detectors are presented.
Furthermore, as the KAGRA Pcal systems are the first systems used in a cryogenic interferometer, the knowledge gained from their investigation will be valuable for future GW detectors employing cryogenic techniques such as the Einstein Telescope (ET) \cite{Punturo_2010}, \cite{Hild_2011} and Cosmic Explorer (CE) \cite{1907.04833}.

\section{Theoretical Framework}
\subsection{Calibration of an interferometric GW detector}
\par
\begin{figure}[htbp]
    \centering
    \includegraphics[width=0.8\textwidth]{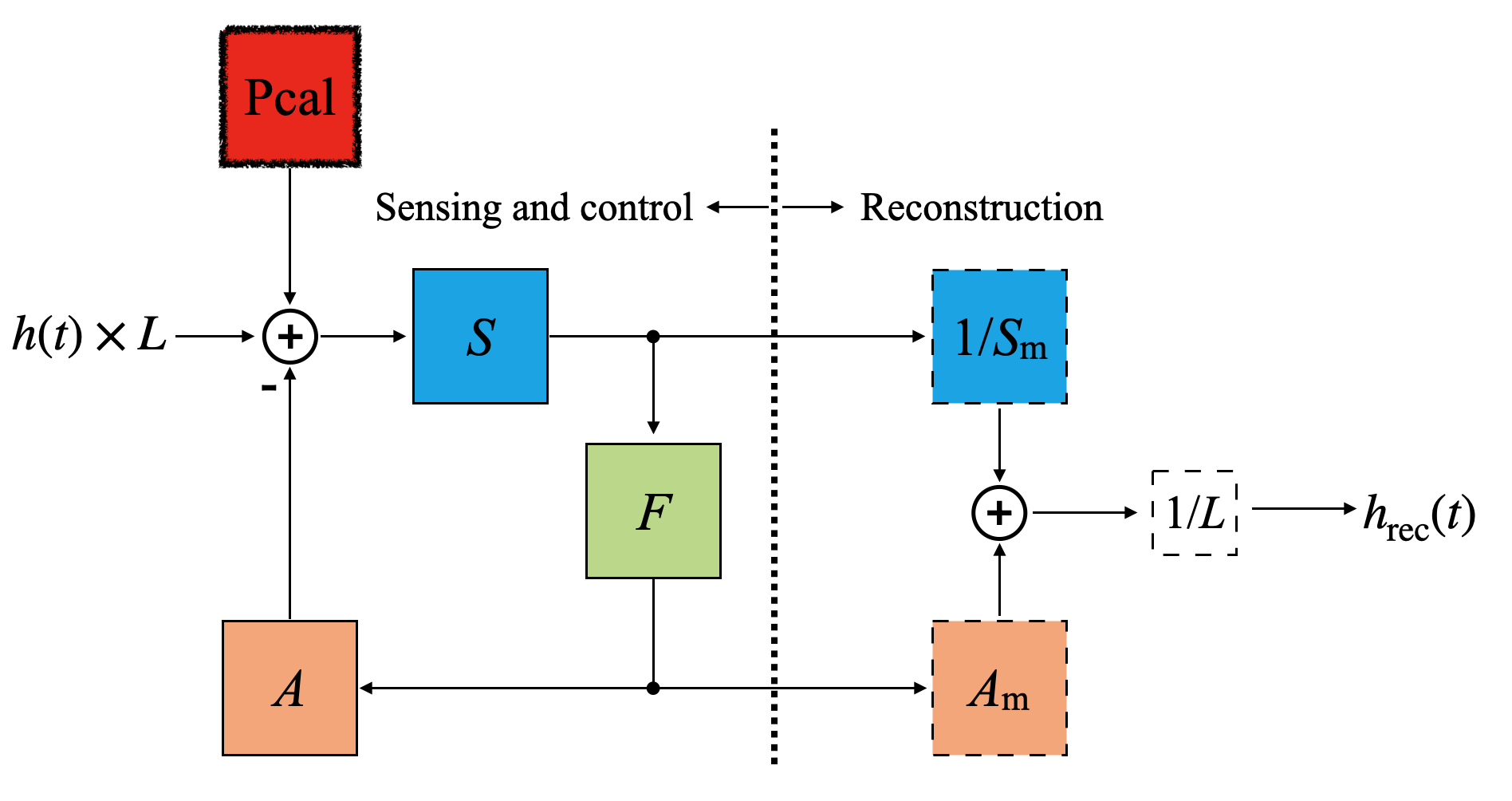}
    \caption{Schematic diagram of the differential arm length (DARM) control loop and strain data $h(t)$ reconstruction. $S$, which is the sensing part of the loop, is an interferometer, and $F$ is a digital control filter. $A$ is the actuation part, which includes the coil magnet actuator and the suspension transfer function. The {\em external} strain data $h_{\rm rec}$ are reconstructed using models of the sensing and actuation parts, $S_{\rm m}$ and $A_{\rm m}$. The photon calibration (Pcal) system serves as a reference signal injector at the point of the gravitational-wave injection.}
    \label{fig:DARM_control_loop_and_reconstruction}
\end{figure}

 The currently operated large laser interferometric GW telescopes mainly consist of six mirrors and a beam splitter (a power recycling mirror, a beam splitter, a signal recycling mirror, two input test mass (ITMX and ITMY) mirrors, and two end test mass (ETMX and ETMY) mirrors).
To maximize the sensitivity of GW detectors to GW strains, the telescopes are complex instruments that rely on high-precision real-time feedback control loops.
The most important control loop is the differential arm (DARM) control loop, which controls and senses the differential arm length.
This loop, shown schematically in Figure \ref{fig:DARM_control_loop_and_reconstruction}, features a Pound–Drever–Hall (PDH) signal (which acts as a sensor ($S$)), a control filter ($F$), and actuators ($A$) \cite{Drever:1983qsr} \cite{Regehr:95}.
As the strain variations $h(t)$ caused by GWs affect this loop, $h(t)$ can be reconstructed from the loop signals using a model of the sensor and actuator responses \cite{Viets_2018}.
However, some parameters are difficult to estimate using only modeling.
Therefore, a system that injects calibrated reference signals into this control loop is required.
Pcal systems, which project auxiliary laser beams onto the end test mass (ETM) to induce calibrated displacements via radiation pressure, are the main calibration systems at KAGRA, LIGO, and Virgo, while the Newtonian calibration (Ncal) system at Virgo also serves as a primary calibration reference \cite{10.1063/5.0147888}, \cite{Aubin_2024}.

\subsection{KAGRA Pcal system}
\par
\begin{figure}[htbp]
    \centering
    \includegraphics[width=0.9\textwidth]{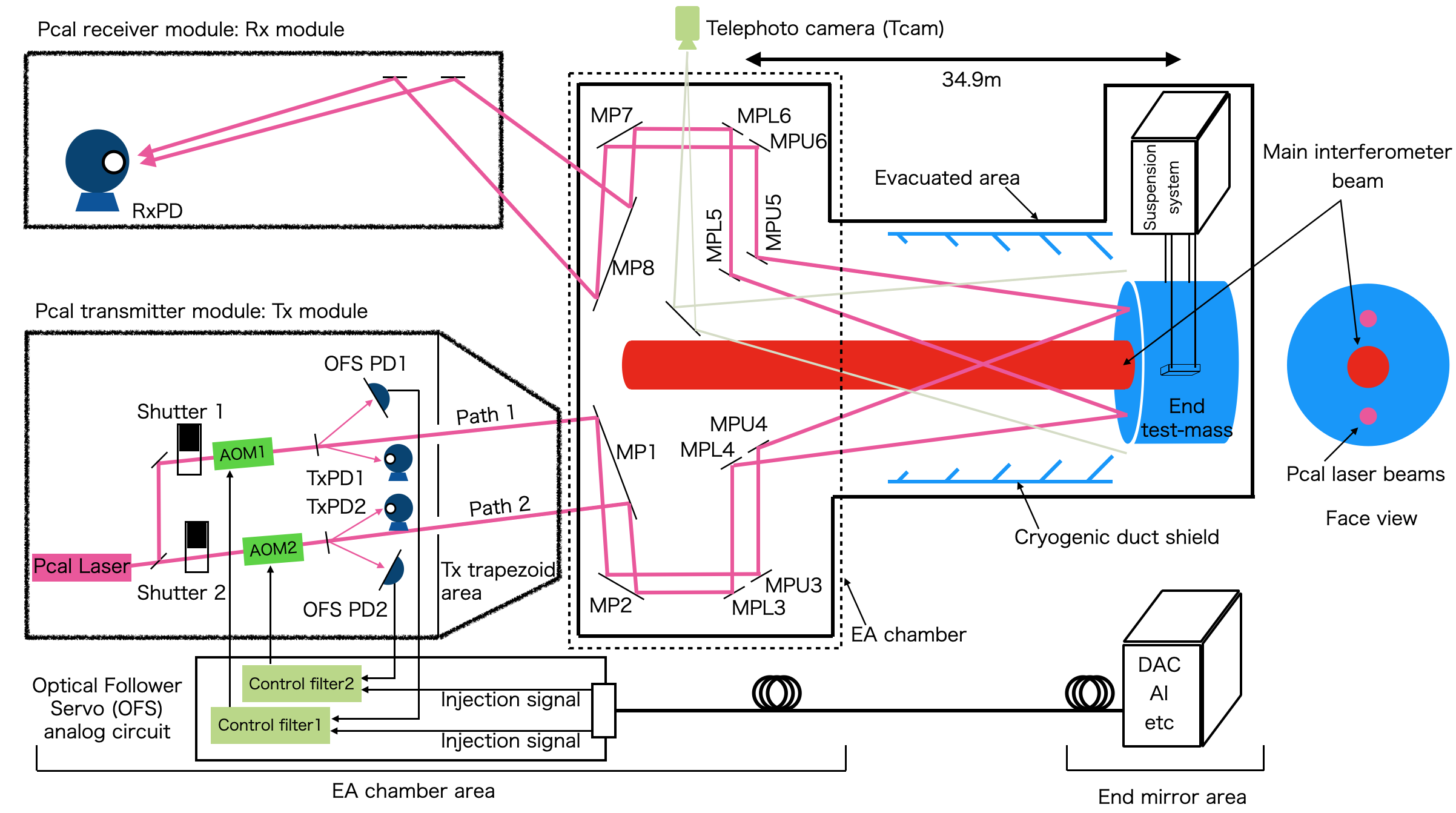}
    \caption{Operating concept of the photon calibration (Pcal) systems in KAGRA. To avoid interference with the cryogenic system, the Pcal system components are located in the EA chamber area. The two Pcal laser beams (path 1 and path 2) generated by the Tx module hit the end test mass (ETM), and they apply reference signals via the radiation pressure to the interferometer; the reflected Pcal beams are received by the Rx module. The Pcal laser beam power on the ETM is estimated from the data measured by RxPD and TxPD, which are integrating sphere type laser power sensors.}
    \label{fig:KAGRA_Pcal_concept}
\end{figure}

 The Pcal concept was initially tested by the Glasgow group \cite{CLUBLEY200185} and was later implemented by the GEO600 group \cite{MOSSAVI20061}.
During O4a, two Pcal systems were operating in KAGRA; the main Pcal-X system, which was used to inject reference signals on ETMX and the Pcal-Y system, which was used as a stand-by Pcal system in the ETMY area.
The operating concepts and components of the two Pcal systems are the same; Figure \ref{fig:KAGRA_Pcal_concept} shows the schematic of a Pcal system.
Each KAGRA Pcal system consists of a transmitter (Tx) module, an end A (EA) chamber, and a receiver (Rx) module.
\par The Tx module generates two stable laser beams (Pcal beams: path 1 and path 2), which are injected into the vacuum area, and imposes power modulation on the beams using acousto-optic modulators (AOMs).
The output beam generated by the laser source (Yb fiber laser provided by KEOPSYS with a central wavelength of 1047\ nm and a maximum output laser power of 20\ W; model CYFL-TERA-20-LP-1047-AM1-RG0-OM1-T305-C1 for Pcal-X and CYFL-TERA-20-LP-1047-AM1-RG0-OM1-B306-C31 for Pcal-Y) is split into two beams using a beam splitter (BS).
Both beams pass through an AOM and a beam sampler before being injected into the vacuum area.
An optical follower servo (OFS) control system composed of the AOM, a photodetector (OFS PD), and electronic circuits, including control filters, is used for each Pcal beam to stabilize the laser power and generate the reference signals with amplitude modulation.
The reference and some control signals, such as the loop gain and DC offset of the OFS loop, are controlled by the KAGRA digital system using digital-to-analog converters and anti-imaging circuits.
\par The two Pcal beams (path 1 at a high point and path 2 at a low point on the ETM) injected into the EA chamber are aligned and injected at appropriate positions on the ETM.
The reason for using two beams instead of one is to avoid elastically deforming the surface of the ETM in the region sensed by the interferometer.
By positioning the beams away from the center and near the nodal circle of the drumhead mode, we minimize the excitation of this mode, which has the highest coupling to the average length sensed by the interferometer beam.
A cryogenic system, which surrounds the ETMs in KAGRA, is used to cool them down to 20 K. The cryogenic duct shield block room temperature radiation emitted from the beam duct to the ETM.
This is one of the reasons for placing the KAGRA Pcal system 34.9 m away from the ETM; one advantage of this configuration is the reduction in the uncertainty of the incident angle measurements; its disadvantage is that it increases the difficulty of Pcal beam alignment.
To overcome this disadvantage, before O4a, we installed picomotors on the mirrors (MPU4 and MPL4) just before the ETM to adjust the beam positions on the ETM \cite{Chen:2023fb}.
We used a camera with a telescope (telephoto camera, Tcam) out of the EA chamber to monitor the ETM surface and take pictures to estimate the beam positions on the ETM.
\par After hitting the ETM, the beams are reflected back to the EA chamber and directed to the Rx module.
Two laser power sensor (TxPD1, 2) installed in the Tx module and one installed in the Rx module (RxPD) are used to measure the transmitted laser power of the Tx module and the laser power received by the Rx module. These powers are important to estimate the reference signal values, as will be explained later.

\par According to previous studies, the displacement (frequency domain) Pcal applies on an interferometer from an ETM is given as follows\cite{10.1063/1.4967303}:
\begin{equation}
    x(f) = -\frac{2P(f)\cos\theta}{cM(2\pi f)^2}\left[ 1+(\vec{a}\cdot\vec{b})\frac{M}{I} \right],
    \label{eq:Pcal_main}
\end{equation}
where $x(f)$ is the ETM displacement due to the Pcal beams, $P(f)$ is the total Pcal laser beam power reflecting from the ETM, $\theta$ is the incident angle of the Pcal beams to the ETM, $c$ is the speed of light, $f$ is the operation frequency, $\vec{a}$ is the power-weighted sum of the displacement vectors (with respect to the center of the face of the ETM) of the positions of the two Pcal beams, $\vec{b}$ is the displacement vector for the main interferometer beam, $M$ is the ETM mass, and $I$ is the ETM inertial moment for yaw and pitch directions.
Since the ETM can be approximated as a right cylinder, $I$ can be written in terms of the radius $r$ and thickness $l$ as follows:
\begin{equation}
    I = \left( \frac{r^2}{4} + \frac{l^2}{12} \right)M.
    \label{eq:Pcal_I}
\end{equation}
\par Considering that each parameter in the above equations contains uncertainty, the Pcal uncertainty, which represents the uncertainty in the KAGRA calibration reference signal, can be obtained by calculating $x(f)$.
The Pcal laser beam power $P$ and the beam positions $\vec{a}, \vec{b}$ on the ETM are the most difficult to estimate and are the largest contributors to the Pcal total uncertainty.
Next, we present the method employed to estimate these parameters and their uncertainties during O4a.

\section{Estimation method for the Pcal laser beam power on ETM}
\par
\begin{figure}[ht]
  \centering
  \includegraphics[width=0.6\textwidth]{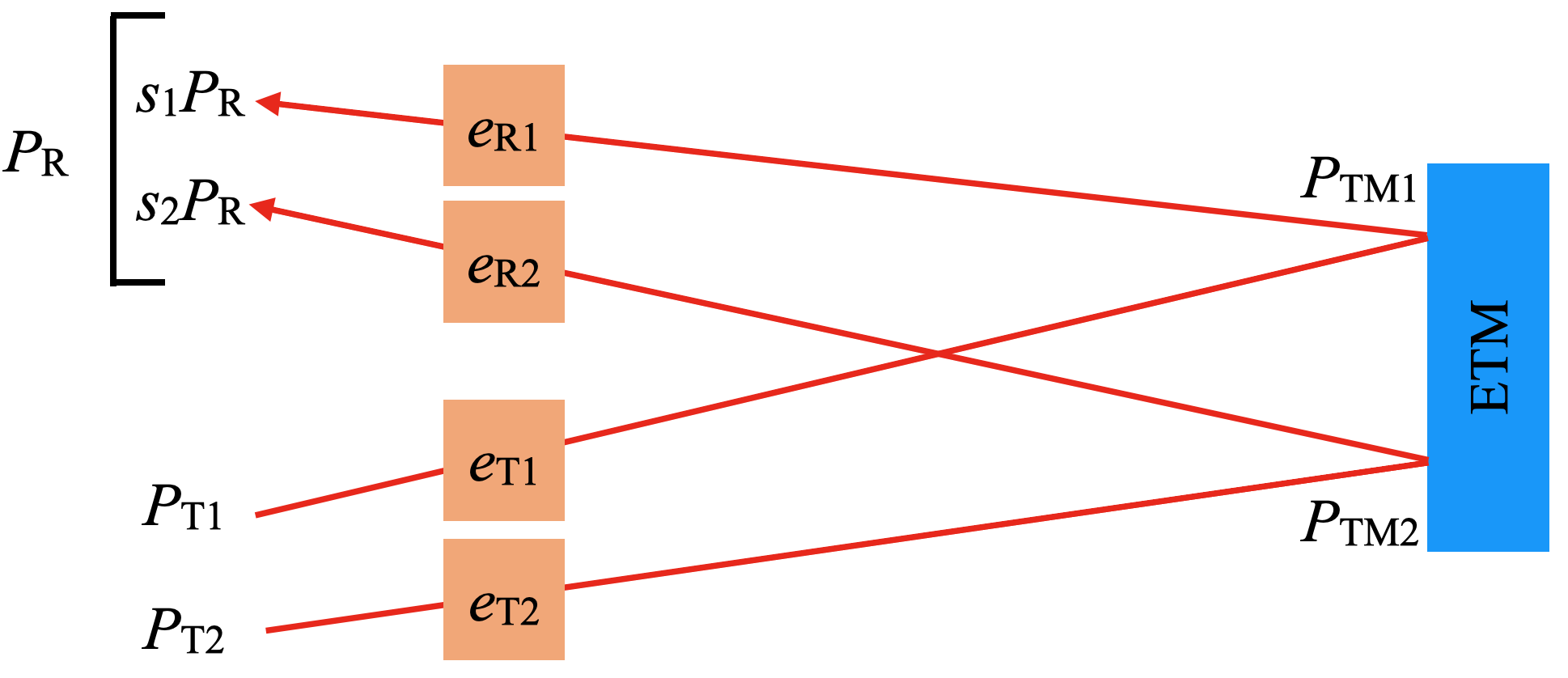}
  \caption{Optical efficiencies and laser power estimation on the end test mass (ETM) using the integrating sphere type laser power sensors.}
  \label{fig:optical_efficiency}
\end{figure}

 We adopted a method to estimate the laser power $P$ on the ETM using all power sensors (two TxPDs and one RxPD).
$P$ can be estimated using the measured optical efficiencies of the Pcal laser beams from both the Tx and Rx modules.
The calculation method for the laser power in each path is presented in Table \ref{table:power_estimation}, where $e_{\rm T1}, e_{\rm T2}, e_{\rm R1}$, and $e_{\rm R2}$ are the optical efficiencies, and $s_1$ and $s_2$ are the separation ratios (Figure \ref{fig:optical_efficiency}).
Direct laser power measurements inside the chambers are not feasible; however, the optical efficiency is expected to be similar between the input and output paths because the optical paths are comparable (e.g., same number of relay mirrors).

Then, $P$ can be estimated as follows:

\begin{equation}
    P = \frac{P_{\rm T1}e_{\rm T1} + s_1e_{\rm R1}^{-1}P_{\rm R} + P_{\rm T2}e_{\rm T2} + s_2e_{\rm R2}^{-1}P_{\rm R}}{2}
    \label{eq:P_TM_estimation}
\end{equation}

\begin{table}[htbp]
    \centering
    \caption{Laser power estimation on the end test mass (ETM)}
    \begin{tabular}{|c|c|c|}
        \hline
         & Path 1 & Path 2 \\
        \hline
        \hline
        Estimation from the Tx side & $P_{\rm TM1} = e_{\rm T1}P_{\rm T1}$ & $P_{\rm TM2} = e_{\rm T2}P_{\rm T2}$ \\
        \hline
        Estimation from the Rx side & $P_{\rm TM1} = s_1e_{\rm R1}^{-1}P_{\rm R}$ & $P_{\rm TM2} = s_2e_{\rm R2}^{-1}P_{\rm R}$ \\
        \hline
    \end{tabular}
    \label{table:power_estimation}
\end{table}

\subsection{Scheme used for the calibration of Pcal power sensors}
\par
\begin{figure}[ht]
  \centering
  \includegraphics[width=0.9\textwidth]{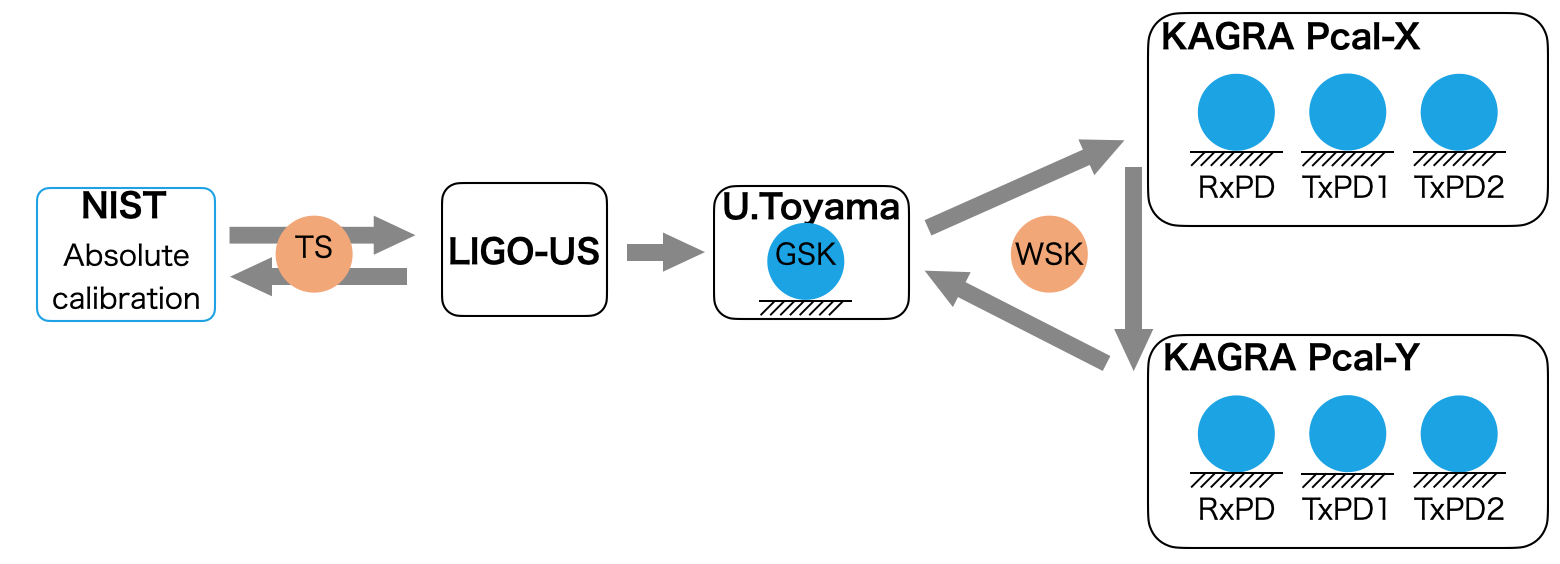}
  \caption{Calibration factor transfer. NIST provided the absolute calibration of a transfer standard to LIGO-US to calibrate the gold standard of KAGRA (GSK) at LIGO-US. The working standard of KAGRA (WSK) was calibrated using the GSK at University of Toyama; then, the power sensors in the photon calibration (Pcal) systems were calibrated using the WSK at the end stations, where the Pcal systems were located. The calibration processes were regularly performed at the KAGRA end stations and University of Toyama.}
  \label{fig:IS_calibration_procedure}
\end{figure}

 The Pcal system laser power sensors incorporate integrating spheres and unbiased InGaAs photodetectors with integrated transimpedance amplifiers.
For the O4 observing run, the KAGRA sensors were calibrated using a {\em gold standard} (GSK) that was provided by LIGO.
Its calibration was provided by a {\em transfer standard} (TS) that was calibrated by the National Institute of Standards and Technology (NIST) in Boulder, Colorado.
Propagation of the calibration to the power sensors operating at the KAGRA end stations is shown schematically in Figure \ref{fig:IS_calibration_procedure}.
We maintained GSK at University of Toyama and used a {\em working standard} (WSK) to transfer the calibration factor to the TxPDs and RxPD at the KAGRA end stations.
We determined the absolute responsivity of the power sensors at the KAGRA end stations through multiple comparative measurements using two different integrating sphere type power sensors.

The responsivity $\rho$ for each power sensor can be derived from the following equations:
\ref{eq:rho_Rx}, \ref{eq:rho_TxPD1}, \ref{eq:rho_TxPD2}.
\begin{eqnarray}
    \rho_{\rm Rx} [{\rm V/W}] &=& \rho_{\rm GSK} \times \alpha_{\rm WSK/GSK} \times \alpha_{\rm RxPD/WSK}, \label{eq:rho_Rx} \\
    \rho_{\rm TxPD1} [{\rm V/W}] &=& \rho_{\rm GSK} \times \alpha_{\rm WSK/GSK} \times \alpha_{\rm TxPD1/WSK}, \label{eq:rho_TxPD1} \\
    \rho_{\rm TxPD2} [{\rm V/W}] &=& \rho_{\rm GSK} \times \alpha_{\rm WSK/GSK} \times \alpha_{\rm TxPD2/WSK}, \label{eq:rho_TxPD2}
\end{eqnarray}
where $\rho$ is the responsivity that converts the output voltage of a power sensor to the laser power, and $\alpha_{\rm A/B}$ is the output voltage ratio when the same laser power is injected into the two integrating spheres A and B. We call this ratio the responsivity ratio of A and B.
$ \rho_{\rm GSK} $ is the calibration, traceable to SI units, provided by NIST and LIGO-US. Together with it’s uncertainty, it is given by

\begin{equation}
    \rho_{\rm GSK} = 4.70095 \pm 0.0048 \ [{\rm V/W}].
    \label{eq:rho_GSK_value}
\end{equation}

\subsection{Calibration of the working standard of KAGRA (WSK)}
\par
\begin{figure}[ht]
  \centering
  \includegraphics[width=0.9\textwidth]{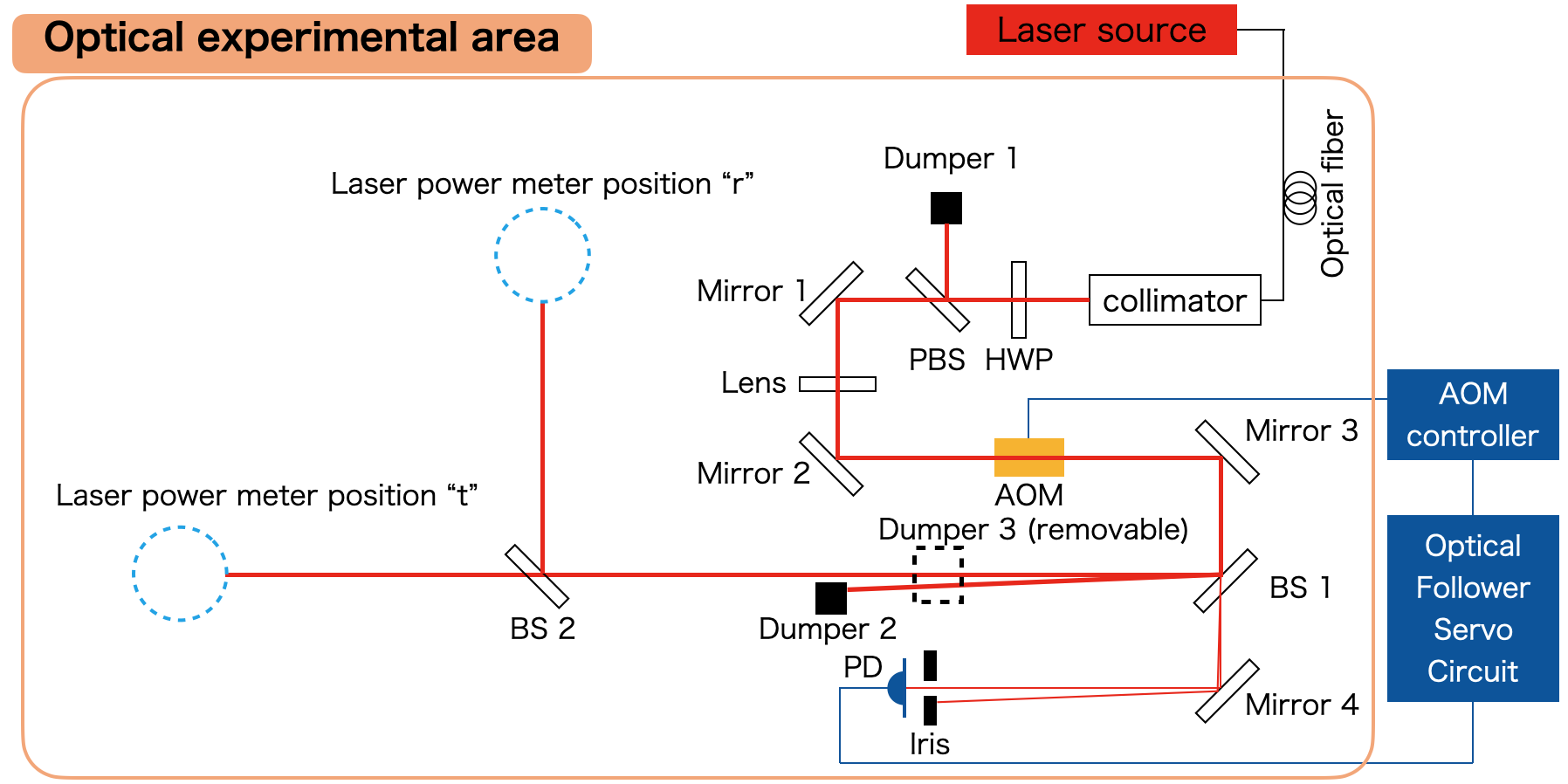}
  \caption{Measurement setup for a comparison measurement between GSK and WSK at University of Toyama. A laser power stabilization system (called the OFS loop) was used for laser power stabilization in the measurement. To reduce the effect of the beam splitter 2 (BS2) reflectance and transmittance imperfections, GSK and WSK were measured at positions "r" and "t" interchangeably.}
  \label{fig:CAL_at_UToyama}
\end{figure}

 Here, we describe the calibration procedure of the responsivity ratio ($\alpha_{\rm WSK/GSK}$) at University of Toyama.
Figure \ref{fig:CAL_at_UToyama} shows this concept.
This system employs a feedback control loop for laser power stabilization referred to as the OFS.
The stabilized laser beam is separated via a beam splitter (BS2) and injected into two power sensors at positions "t" and "r".
We placed GSK and WSK at positions "t" and "r" to perform a comparison measurement, which suppresses the effects of laser power fluctuation and the nonideal operation of BS.

\par As the WSK–GSK calibration procedure, we performed the following steps: 1) We fixed WSK and GSK at positions "r" and "t", respectively (i.e., at the reflected and transmitted ports, respectively) of BS2; then, we turned ON the laser and the OFS loop and removed dumper 3. 2) After a certain equipment warming-up time ($\sim$1h), we measured the dark noise by putting back into place damper 3 (220 data points in $\sim 1$ min; $\bm{V}_{i, {\rm rWSK, dark}}, \bm{V}_{i, {\rm tGSK, dark}}$). 3) We removed dumper 3 and measured the main data (1100 data points in $\sim5$ min; $\bm{V}_{i, {\rm rWSK}}, \bm{V}_{i, {\rm tGSK}}$). 4) We swapped the positions of GSK and WSK (i.e., WSK was placed at position "t", and GSK was placed at position "r") and measured the dark noise again using dumper 3 (220 data points in $\sim1$ min; $\bm{V}_{i, {\rm tWSK, dark}}, \bm{V}_{i, {\rm rGSK, dark}}$). Finally, we removed dumper 3 and measured the main data (1100 data points in $\sim5$ min; $\bm{V}_{i, {\rm tWSK}}, \bm{V}_{i, {\rm rGSK}}$). We repeated this sequence of measurements five times in a row ($i = 1, ..., 5$).

\par We set the laser power to 3 W and controlled the OFS loop to inject $\sim380$ mW into each power sensor.
We used an ADCMT 7352A Digital Multimeter, which can simultaneously measure two signals, for data acquisition.
The optical experimental area was covered by a black anodized aluminum metal box to reduce the impact of the external optical noise.

\par The method used to calculate $\alpha_{\rm WSK/GSK}$ is described below.
Given that $t$ and $r$ are the transmittance and reflectance, respectively, of BS2, $\alpha_{\rm A/B}$ can be calculated as follows:
\begin{eqnarray}
    \alpha_{\rm A/B} &=& \sqrt{\frac{V_{\rm tA} \times V_{\rm rA}}{V_{\rm tB} \times V_{\rm rB}}} = \sqrt{\frac{\rho_{\rm A} t P \times \rho_{\rm A} r P}{\rho_{\rm B} t P \times \rho_{\rm B} r P}} \\
    &=& \frac{\rho_{\rm A}}{\rho_{\rm B}},
    \label{eq:alpha_ApB}
\end{eqnarray}
where $\rho_{\rm A}$ and $\rho_{\rm B}$ are the responsivities, which serve as the conversion factors from the laser power to the output voltage of power sensors A and B.
If the transmittance and reflectance of BS2 are stable, this method can eliminate their impact.
The following formula, which incorporates dark noise removal, the contemporaneous ratio calculation, and averaging, was used to calculate $\alpha_{\rm WSK/GSK}$ in each experiment.
\begin{eqnarray}
    \alpha_{i, {\rm WSK/GSK}} &=& \sqrt{\left\langle \frac{\bm{V}_{i, {\rm rWSK}} - \langle \bm{V}_{i, {\rm rWSK, dark}} \rangle}{\bm{V}_{i, {\rm tGSK}} - \overline{\langle \bm{V}_{i, {\rm tGSK, dark}} \rangle}} \right\rangle \left\langle \frac{\bm{V}_{i, {\rm tWSK}} - \langle \bm{V}_{i, {\rm tWSK, dark}} \rangle}{\bm{V}_{i, {\rm rGSK}} - \langle \bm{V}_{i, {\rm rGSK, dark}} \rangle} \right\rangle} \\
    \alpha_{\rm WSK/GSK} &=& \langle [ \alpha_{1, {\rm WSK/GSK}}, ... {\alpha_{5, \rm WSK/GSK}} ] \rangle
    \label{eq:alpha_WSKpGSK}
\end{eqnarray}
Here, $\langle \bm{A} \rangle$ denotes the weighted average of the elements of a vector $\bm{A}$, calculated using the method outlined in \ref{appendix:average}. The uncertainty associated with the weighted average is computed as the standard deviation (SD) as described in the same section.
Furthermore, $\bm{A}/\bm{B}$ denotes that all ratios between the components of vectors $\bm{A}$ and $\bm{B}$ are calculated separately, and then the vector of the ratios is obtained.

\par We conducted this measurement roughly once a month to accumulate long-term data for a six-month period before O4a started.

\subsection{Calibration of TxPD and RxPD}

\par In this subsection, we present the calculation of the responsivity ratios ($\alpha_{\rm RxPD/WSK}$, $\alpha_{\rm TxPD1/WSK}$, and $\alpha_{\rm TxPD1/WSK}$), the optical efficiencies $e_1$ and $e_2$, and the separation ratios $s_1$ and $s_2$, which are required to calculate the laser power on the ETM, using Equation (\ref{eq:P_TM_estimation}).

\par To estimate $\alpha$ for TxPDs, we placed WSK at the trapezoid area of the Tx module to capture a Pcal laser beam just before leaving the Tx module and measured the output voltage ratio between TxPD and WSK:
\begin{equation}
    \alpha_{\rm TxPD/WSK} = V_{\rm TxPD}/V_{\rm WSK}.
    \label{eq:alpha_TxPDpWSK}
\end{equation}

\par $\alpha$ for RxPD was estimated by measuring the output voltage of RxPD and placing WSK in the position of RxPD; then, the ratio of the RxPD output and WSK was calculated.
During these measurements, the TxPD output was always monitored and was used to normalize the RxPD and WSK data.
This measurement was performed with each of the two Pcal beam paths.
In the following equations, $j = 1, 2$ indicates path 1 or path 2.
\begin{equation}
    \alpha_{\rm RxPD/WSK} = \frac{V_{\rm RxPD}}{V_{{\rm TxPD}j}} \left/ \frac{V_{\rm WSK}}{V_{{\rm TxPD}j}} \right.
    \label{eq:alpha_RxPDpWSK}
\end{equation}

\par The measured optical efficiency $e_j$ was obtained from the ratio of the WSK outputs in the Tx and Rx modules.
Again, the output of the TxPDs was used as a reference for the input laser power and measured for paths 1 and 2 independently of each other.
The optical efficiencies of path 1(2) at the transmitter(receiver) side are $e_{\rm T1}, e_{\rm R1}, e_{\rm T2}$, and $e_{\rm R2}$, which are expressed along with $e_j$ as follows:
\begin{eqnarray}
    e_j &=& \frac{V_{\rm WSK, Rx}}{V_{{\rm TxPD}j}} \left/ \frac{V_{\rm WSK, Tx}}{V_{{\rm TxPD}j}} \right. \label{eq:e} \\
    e_{{\rm T}j} &=& e_{{\rm R}j} = \frac{1+e_j}{2} \pm \frac{1-e_j}{2\sqrt{3}} \label{eq:e_TR}
\end{eqnarray}
Here, since it is unclear whether the optical power loss occurs between the Tx module and the ETM or between the ETM and the Rx module, we expect $e_{{\rm T}j}$ and $e_{{\rm R}j}$ to be between 1 and $e_j$.
Thus, we regarded these optical efficiencies as "Type-B" measurement values, as categorized by NIST \cite{typeB-NIST}.

\par To estimate the separation ratio $s$, we measured the RxPD and TxPD outputs for each beam, and calculated the laser power in RxPD normalized by the laser power at ETM using $e_j$:
\begin{equation}
    P_j = \frac{V_{{\rm RxPD}, j}}{\rho_{\rm RxPD}} \left/ \left(\frac{V_{{\rm TxPD}j}}{\rho_{{\rm TxPD}j}}e_{{\rm T}j}\right) \right. \label{eq:s_P_j} .
\end{equation}
Then, the separation ratios were calculated as follows:\begin{equation}
    s_j = \frac{P_j}{P_1 + P_2}
    \label{eq:separation_ratio}	
\end{equation}

\par The actual calibration procedure for the RxPD and TxPDs at the end stations involves three main steps: (A) measurement using the RxPD for the calibration of RxPD, (B) measurement using WSK at the Rx module for the calibration of RxPD, and (C) measurement using TxPDs and WSK at the Tx module for the calibration of TxPDs.
\par In step (A), we placed a damper at the Tx trapezoid in path 2 to ensure that only path 1 reaches the RxPD from the Tx module via the ETM, and recorded the outputs ($\bm{V}_{\rm A, Rx, p1}$ and $\bm{V}_{\rm A, Tx1, p1}$) from RxPD and TxPD1. Next, we switched the damper position from path 2 to path 1 to ensure that only path 2 reaches the RxPD from the Tx module, and recorded the outputs ($\bm{V}_{\rm A, Rx, p2}$ and $\bm{V}_{\rm A, Tx2, p2}$) from RxPD and TxPD2.
\par In step (B), we replaced RxPD with WSK and measured the WSK, TxPD1, and TxPD2 outputs ($\bm{V}_{\rm B, WSK, p2}, \bm{V}_{\rm B, Tx2, p2}$, and $\bm{V}_{\rm B, WSK, p2}, \bm{V}_{\rm B, Tx2, p2}$) with path 1 and path 2, respectively, as in step (A).
\par In step (C), we moved WSK to the Tx trapezoid area to measure the output laser beam power from the Tx module. Initially, we placed WSK at path 1 and damped path 2; next, we measured the TxPD1 and WSK outputs ($\bm{V}_{\rm C, WSK, p1}$ and $\bm{V}_{\rm C, Tx1, p1}$). Then, we switched to path 2 and measured $\bm{V}_{\rm C, WSK, p2}$ and $\bm{V}_{\rm C, Tx2, p2}$.
\par Between the main measurements described above, we conducted dark offset measurements for each integrating sphere using shutters 1 and 2 to obtain $\bm{V}_{\rm Tx1, d}, \bm{V}_{\rm Tx2, d}, \bm{V}_{\rm RxPD, d}$, and $\bm{V}_{\rm WSK, d}$.
Additionally, during the calibration of WSK at the end stations, we measured the voltage drop from the WSK output injection port in the Pcal system to the KAGRA digital system using the HIOKI SS7012 reference voltage generator, which has a relative accuracy of 0.02\ \%. The measured drop was approximately 0.18\ \%, so we applied a correction factor of 1.0018 to the WSK readout values to compensate for this difference.
\par During the measurements, the OFS loops were closed to provide power-stabilized laser beams for paths 1 and 2, except for the dark offset measurements. The typical laser power transmitted to ETM was 1\ W in each path.
The KAGRA data acquisition system was used to record all data. In each measurement, the data were recorded for more than 5 min using a 512-Hz sampling frequency. 
\par We used these data to calculate the responsivity ratios $\alpha_{\rm RxPD/WSK}$, $\alpha_{{\rm TxPD}j/{\rm WSK}}$, optical efficiencies $e_j$, and separation ratios $s_j$ described above considering the dark offset measurements of sensors.
\par The following equations were used to calculate $\alpha_{\rm RxPD/WSK}$:
\begin{eqnarray}
    \alpha_{{\rm RxPD/WSK, p}j} &=& \frac{\langle \bm{R}_{{\rm Rx, p}j} \rangle}{\langle \bm{R}_{{\rm WSK, p}j} \rangle}, \\
    \bm{R}_{{\rm Rx, p}j} &=& \frac{\bm{V}_{{\rm A, Rx, p}j} - V_{\rm RxPD, d}}{\bm{V}_{{\rm A, Tx}j, {\rm p}j} - V_{{\rm Tx}j, {\rm d}}}, \\
    \bm{R}_{{\rm WSK, p}j} &=& \frac{\bm{V}_{{\rm B, WSK, p}j} - V_{\rm WSK, d}}{\bm{V}_{{\rm B, Tx}j, {\rm p}j} - V_{{\rm Tx}j, {\rm d}}}
\end{eqnarray}
As shown here, we obtained two values for $\alpha_{\rm RxPD/WSK}$ from one calibration measurement.
As we will discuss later, we determined $\alpha_{\rm RxPD/WSK}$ using all data (from both paths) obtained during a six-month period before the observation run and one measurement just after the run.

\par $\alpha_{\rm TxPD1/WSK}$ and $\alpha_{\rm TxPD2/WSK}$ were calculated using the data measured in step (C) as follows:

\begin{equation}
    \alpha_{{\rm TxPD}j/{\rm WSK}} = \left\langle \frac{\bm{V}_{{\rm C, Tx}j, {\rm p}j} - V_{{\rm Tx}j, {\rm d}}}{\bm{V}_{{\rm C, WSK, p}j} - V_{\rm WSK, d}} \right\rangle
\end{equation}

\par $e_j$ was calculated as follows:

\begin{eqnarray}
    e_j &=& \frac{\langle \bm{R}_{{\rm WSK, Rx, p}j} \rangle}{\langle \bm{R}_{{\rm WSK, Tx, p}j} \rangle}, \\
    \bm{R}_{{\rm WSK, Rx, p}j} &=& \frac{\bm{V}_{{\rm B, WSK, p}j} - V_{\rm WSK, d}}{\bm{V}_{{\rm B, Tx}j, {\rm p}j} - V_{{\rm Tx}j, {\rm d}}}, \\
    \bm{R}_{{\rm WSK, Tx, p}j} &=& \frac{\bm{V}_{{\rm C, WSK, p}j} - V_{\rm WSK, d}}{\bm{V}_{{\rm C, Tx}j, {\rm p}j} - V_{{\rm Tx}j, {\rm d}}}
\end{eqnarray}
The optical efficiencies of the Tx and Rx sides were estimated using Equation \ref{eq:e_TR}.
\par Finally, $s_j$ was calculated using the measured data as follows:
According to Equation \ref{eq:s_P_j}, $P_j$ can be expressed as follows:
\begin{equation}
    P_{j} = \left\langle \frac{\bm{V}_{{\rm A, Rx, p}j} - V_{\rm RxPD, d}}{\bm{V}_{{\rm A, Tx}j} - V_{{\rm Tx}j, {\rm d}}} \right\rangle
    \frac{\alpha_{{\rm TxPD}j/{\rm WSK}}\alpha_{\rm WSK/GSK}\rho_{\rm GSK}}{\alpha_{\rm RxPD/WSK}\alpha_{\rm WSK/GSK}\rho_{\rm GSK}}
    \frac{1}{e_{{\rm T}j}},
\end{equation}
therefore, the separation ratios are 
\begin{eqnarray}
    s_1 &=& \left( 1+\frac{\left\langle \frac{\bm{V}_{\rm A, Rx, p2} - V_{\rm RxPD, d}}{\bm{V}_{\rm A, Tx2, p2} - V_{\rm Tx2, d}} \right\rangle}{\left\langle \frac{\bm{V}_{\rm A, Rx, p1} - V_{\rm RxPD, d}}{\bm{V}_{\rm A, Tx1, p1} - V_{\rm Tx1, d}} \right\rangle}
    \frac{\alpha_{\rm TxPD2/WSK}}{\alpha_{\rm TxPD1/WSK}}
    \frac{e_{\rm T1}}{e_{\rm T2}} \right)^{-1}, \\
    s_2 &=& \left( 1+\frac{\left\langle \frac{\bm{V}_{\rm A, Rx, p1} - V_{\rm RxPD, d}}{\bm{V}_{\rm A, Tx1, p1} - V_{\rm Tx1, d}} \right\rangle}{\left\langle \frac{\bm{V}_{\rm A, Rx, p2} - V_{\rm RxPD, d}}{\bm{V}_{\rm A, Tx2, p2} - V_{\rm Tx2, d}} \right\rangle}
    \frac{\alpha_{\rm TxPD1/WSK}}{\alpha_{\rm TxPD2/WSK}}
    \frac{e_{\rm T2}}{e_{\rm T1}} \right)^{-1}
\end{eqnarray}
\par We performed the above end-station calibration procedure seven times (from Jan. 12, 2023, to June 22, 2023), which includes the KAGRA O4a observation run (from May 25, 2023, to June 21, 2023).
The parameters that should be used for the $h(t)$ reconstruction were calculated using all the data measured during the seven-time calibration.

\section{Estimation of the Pcal laser power on ETM (parameters obtained for O4a)}
\par In this section, we present the measurement results of the WSK calibrations at our Pcal lab at University of Toyama and the TxPD and RxPD calibration results at the end stations in KAGRA.
Considering that the Pcal-X was used during O4a as the reference signal generator in the interferometer, we present only the Pcal-X-related measured results.
\par The measurements were performed seven times between January and June 2023, which included the KAGRA observation period; during this period, we investigated the long-time stabilities and estimated the uncertainties.

\subsection{Calibration of the working standard of KAGRA (WSK)}
\begin{figure}[h!]
  \centering
  \includegraphics[width=0.45\textwidth]{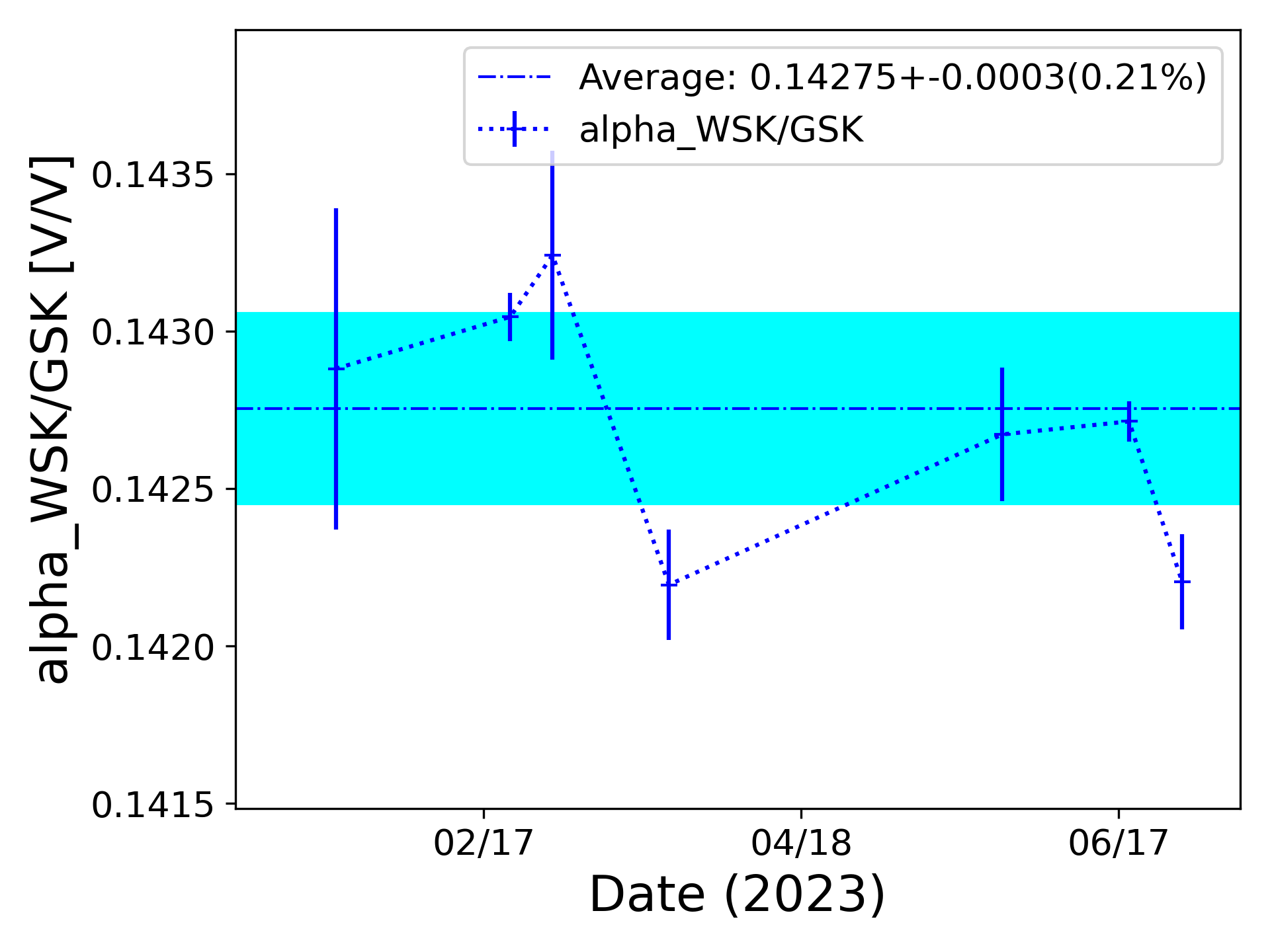}
  \caption{Comparison of WSK and GSK calibration results at University of Toyama ($\alpha_{\rm WSK/GSK}$)}
  \label{fig:result:WSK_over_GSK}
\end{figure}

Figure \ref{fig:result:WSK_over_GSK} shows the $\alpha_{\rm WSK/GSK}$ measured results; the estimated relative uncertainty was 0.21\ \%, which is the weighted SD of the measured data considering individual measurement errors. We calculated this average and uncertainty values following the method outlined in \ref{appendix:average}.
\par Due to the potential presence of unidentified systematic errors, we adopted the standard deviation rather than the standard error with the $1/\sqrt{N}$ factor when estimating uncertainties.
The specific formulas used to calculate both the weighted average and its associated uncertainty are detailed in \ref{appendix:average}.
This approach was similarly applied to the other parameters described in this section.

\subsection{Calibration of TxPDs and RxPD in Pcal-X}
\par In this subsection, we present the measurement results of Pcal-X, which was used to inject the reference signals in O4a. 
The parameters used to estimate the injected Pcal power on the ETMX were $\alpha_{\rm RxPD/WSK}, \alpha_{\rm TxPD1/WSK}, \alpha_{\rm TxPD2/WSK}$, $e_1$, $e_2$, $s_1$, and $s_2$.

\begin{figure}[h!]
  \centering
  \includegraphics[width=0.45\textwidth]{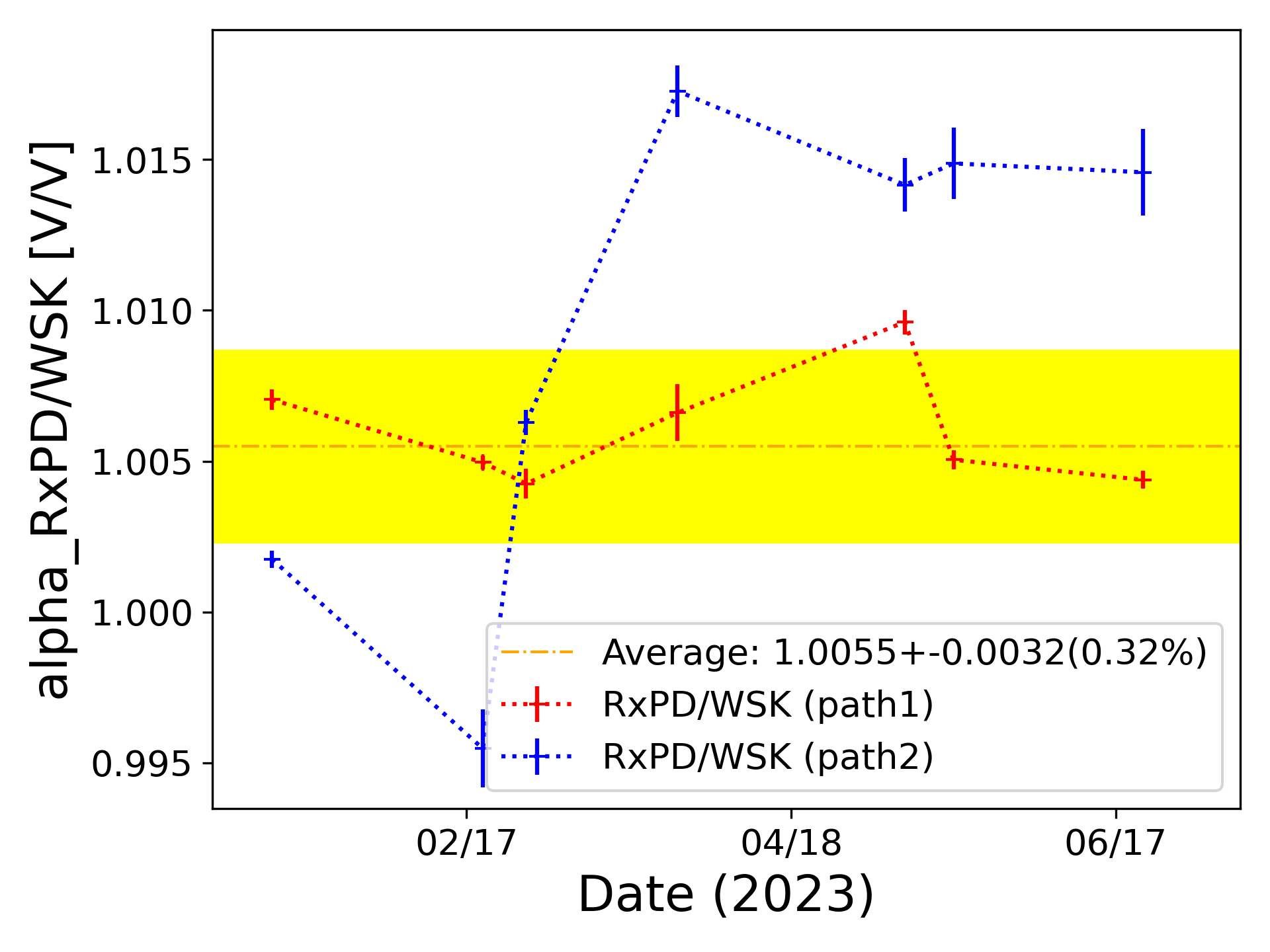}
  \caption{Measured responsivity ratio between Pcal-X RxPD and WSK ($\alpha_{\rm RxPD/WSK}$). We used the data obtained from both paths 1 and 2 to estimate the mean value.}
  \label{fig:result:XPcal_both_path_RxPD_over_WSK}
\end{figure}

\begin{figure}[h!]
    \centering
    \begin{subfigure}[b]{0.45\textwidth}
        \includegraphics[width=\textwidth]{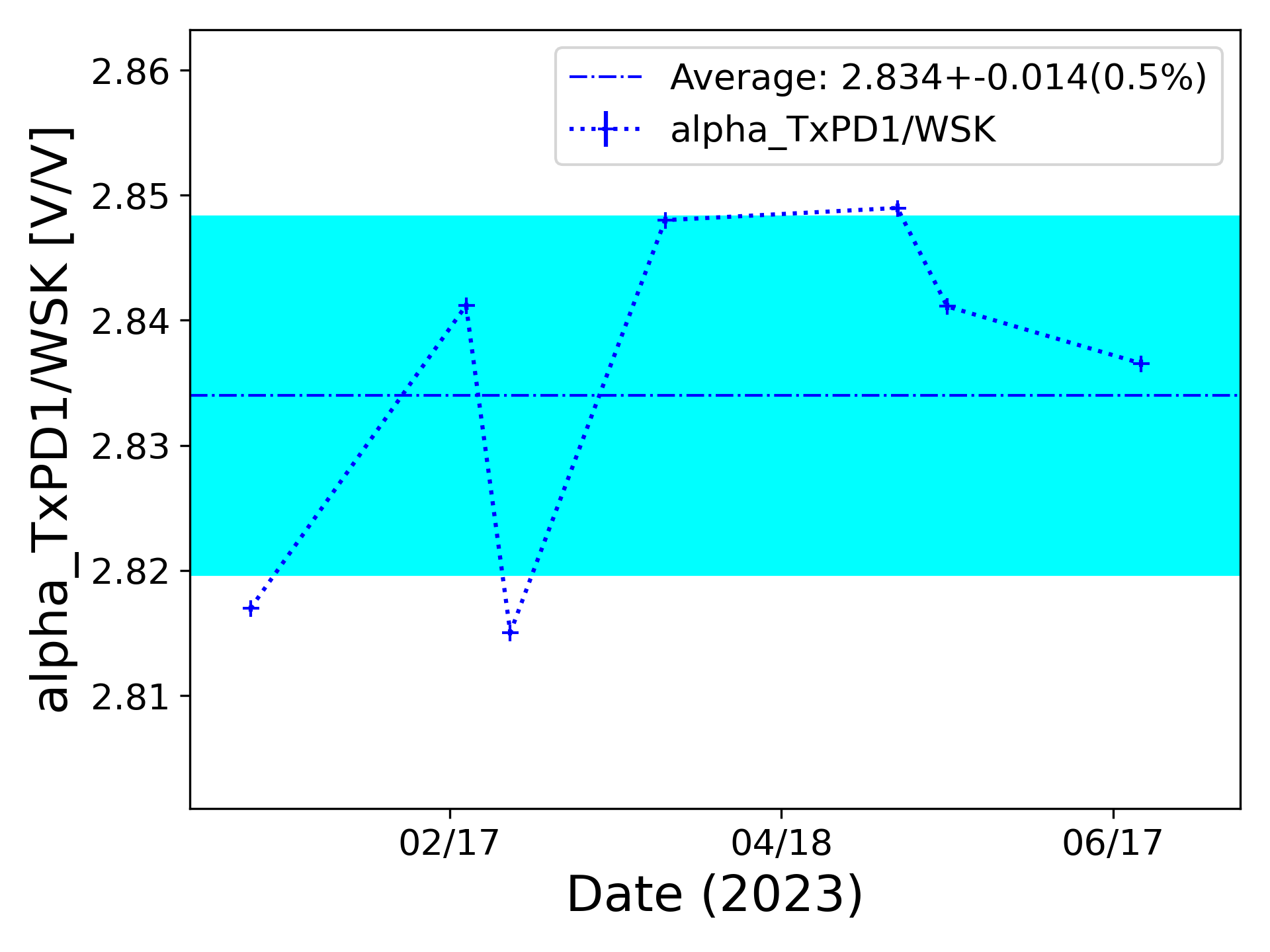}
        \caption{Measured responsivity ratio between Pcal-X TxPD1 and WSK ($\alpha_{\rm TxPD1/WSK}$).}
        \label{fig:result:XPcal_TxPD1_over_WSK}
    \end{subfigure}
    \hfill
    \begin{subfigure}[b]{0.45\textwidth}
        \includegraphics[width=\textwidth]{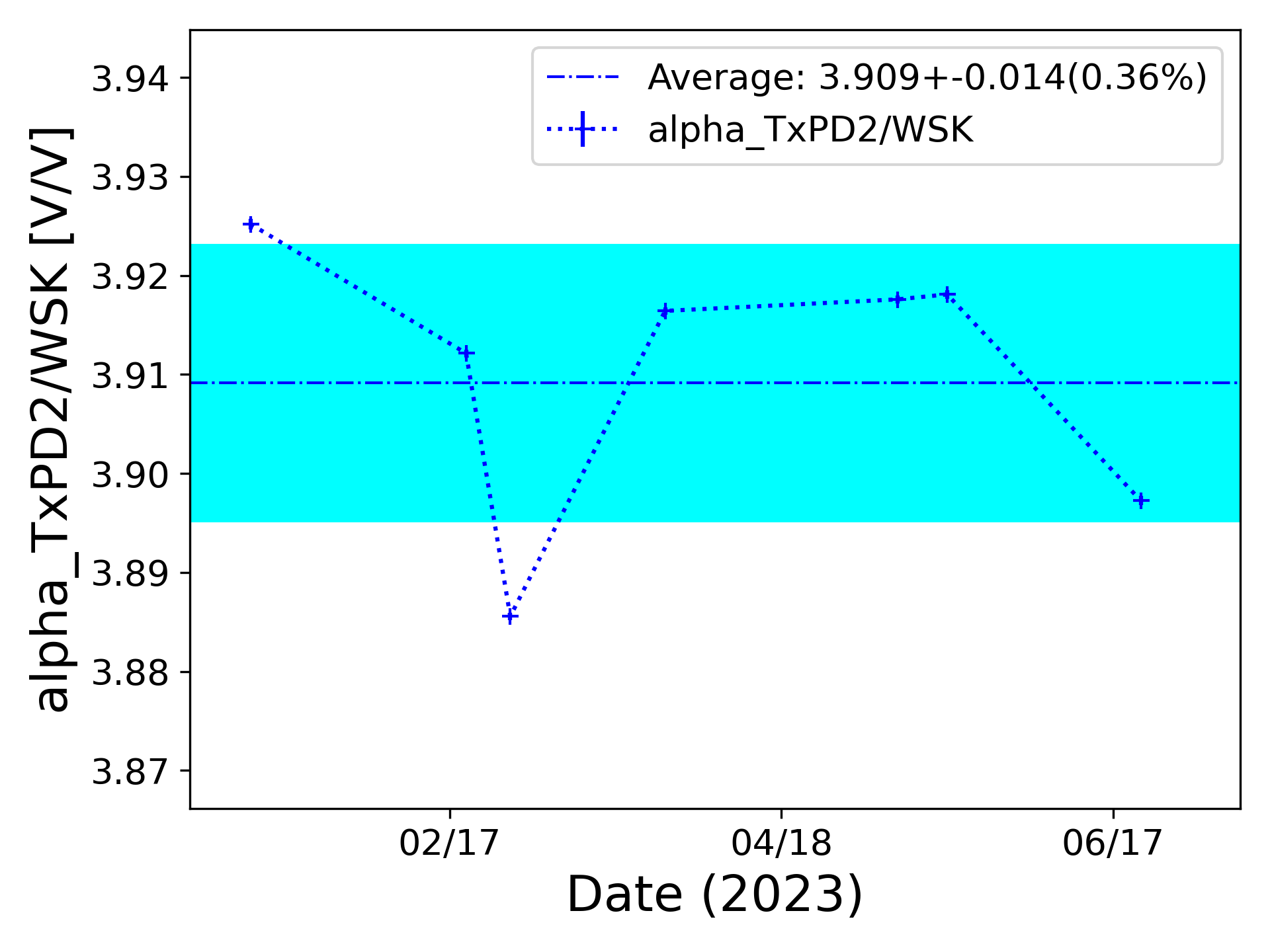}
        \caption{Measured responsivity ratio between Pcal-X TxPD2 and WSK ($\alpha_{\rm TxPD2/WSK}$).}
        \label{fig:result:XPcal_TxPD2_over_WSK}
    \end{subfigure}
    \caption{Measured responsivity ratios of TxPDs based on WSK.}
    \label{fig:reuslt:Pcal-X:alpha_Tx}
\end{figure}

\begin{figure}[h!]
    \centering
    \begin{subfigure}[b]{0.45\textwidth}
        \includegraphics[width=\textwidth]{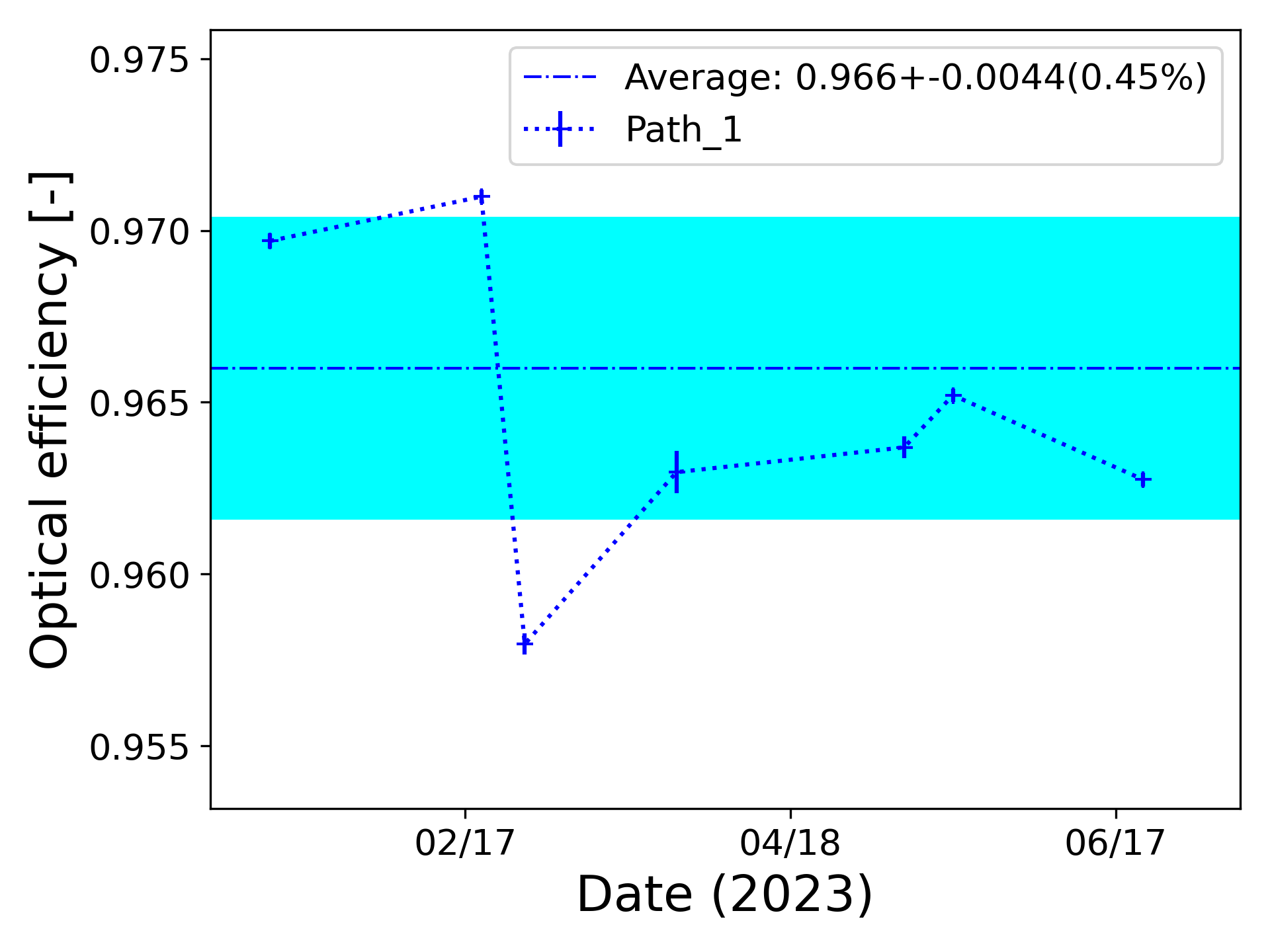}
        \caption{Measured optical efficiency of Pcal-X path 1 ($e_1$).}
        \label{fig:result:XPcal_Path_1_optical_efficiency}
    \end{subfigure}
    \hfill
    \begin{subfigure}[b]{0.45\textwidth}
        \includegraphics[width=\textwidth]{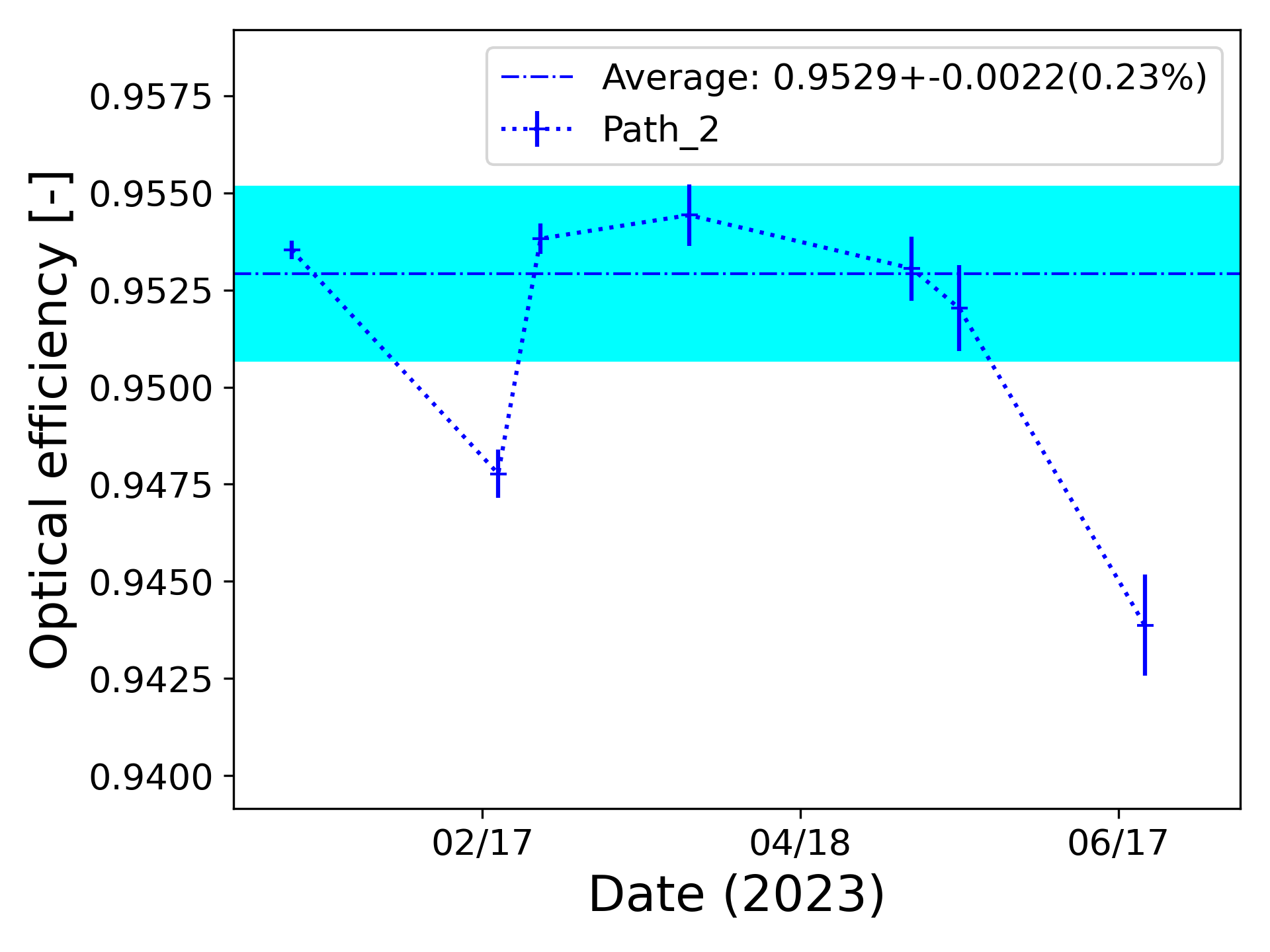}
        \caption{Measured optical efficiency of Pcal-X path 2 ($e_2$).}
        \label{fig:result:XPcal_Path_2_optical_efficiency}
    \end{subfigure}
    \caption{Measured optical efficiencies of the photon calibration (Pcal)-X laser beam paths from the Tx module output to the RxPD position in the Rx module to account for the optical losses primarily caused by optics-like windows and mirrors in the EA vacuum chamber. Since it is unclear whether the loss occurred upstream or downstream of the end test mass (ETM) mirror, the transmitter/receiver-side efficiencies were estimated using the NIST "Type-B" method \cite{typeB-NIST}.}
    \label{fig:reuslt:Pcal-X:optical_efficiency}
\end{figure}

\begin{figure}[h!]
    \centering
    \begin{subfigure}[b]{0.45\textwidth}
        \includegraphics[width=\textwidth]{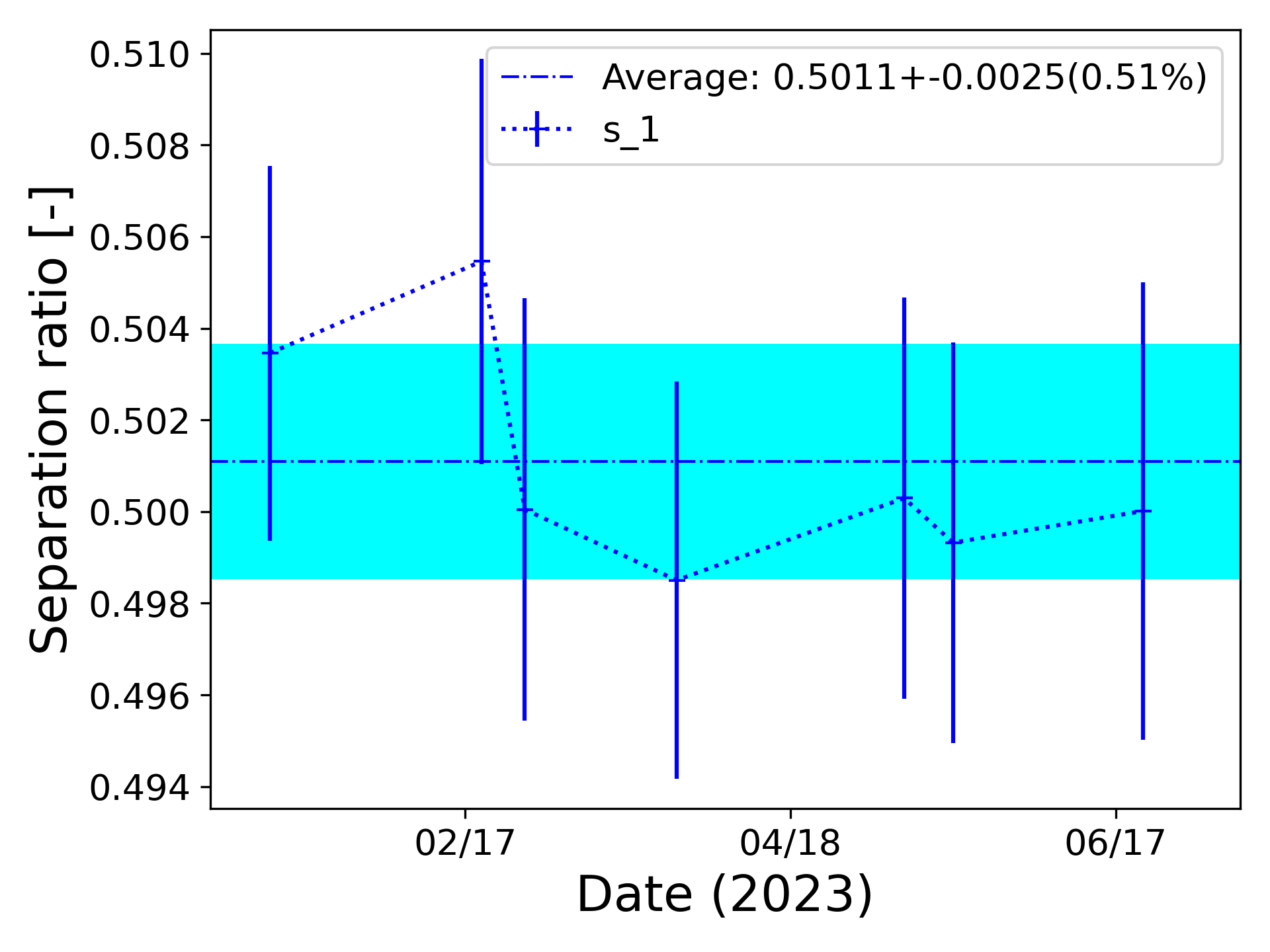}
        \caption{Measured normalized separation ratio of Pcal-X path 1 ($s_1$).}
        \label{fig:result:XPcal_SP_1_separation_ratio}
    \end{subfigure}
    \hfill
    \begin{subfigure}[b]{0.45\textwidth}
        \includegraphics[width=\textwidth]{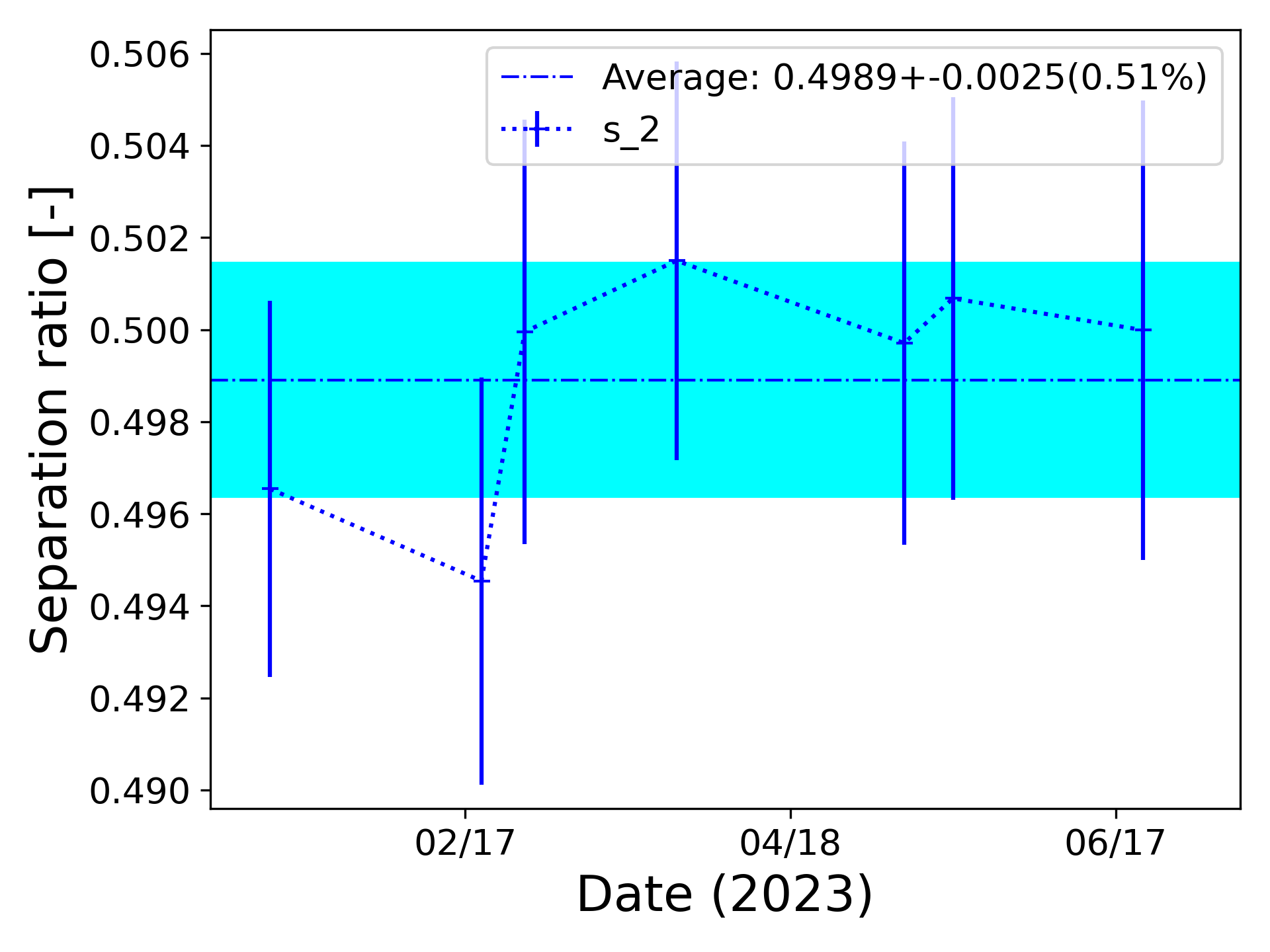}
        \caption{Measured normalized separation ratio of Pcal-X path 2 ($s_2$).}
        \label{fig:result:XPcal_SP_2_separation_ratio}
    \end{subfigure}
    \caption{Measured normalized separation ratios of Pcal-X. As the optical efficiencies for paths 1 and 2 are different, these ratios were not exactly equal to 0.5.}
    \label{fig:result:Pcal-X:separation_ratio}
\end{figure}
\par The data variation (although very small, $\sim$subpercent) was greater than the corresponding error bars.

\subsection{Summary of the integrating sphere calibration}
\par In this subsection, we summarize the measured results. The uncertainty in all measured parameters was less than 0.6\ \%, indicating good parameter stabilization at this level for a six-month period.

\begin{table}[h!]
    \centering
    \begin{tabular}{|l|c|c|}
        \hline
        \textbf{Parameter name} & \textbf{Value and uncertainty} & \textbf{Relative uncertainty [\%]} \\ \hline
        $\alpha_{\rm WSK/GSK}$ [V/V] & $0.14275 \pm 0.00030$ & 0.21 \\ \hline
        $\alpha_{\rm TxPD1/WSK}$ [V/V] & $2.834 \pm 0.014$ & 0.50 \\
        $\alpha_{\rm TxPD2/WSK}$ [V/V] & $3.909 \pm 0.014$ & 0.36 \\
        $e_1$ [-] & $0.9660 \pm 0.0044$ & 0.45 \\
        $e_2$ [-] & $0.9529 \pm 0.0022$ & 0.23 \\
        $\alpha_{\rm RxPD/WSK}$ [V/V] & $1.0055 \pm 0.0032$ & 0.32 \\
        $s_1$ [-] & $0.5011 \pm 0.0025$ & 0.51 \\
        $s_2$ [-] & $0.4989 \pm 0.0025$ & 0.51 \\
        \hline
    \end{tabular}
    \caption{Integrating sphere calibration summary (Pcal-X).}
    \label{tab:IS_calibration_summary}
\end{table}

\newpage

\section{Beam-position estimation method}
\par A telephoto camera (Tcam) system was used to estimate the positions ($\vec{a}_1$ and $\vec{a}_2$) of the Pcal beams and the position ($\vec{b}$) of the main interferometer beam. The Tcam system consisted of a CMOS camera (ASI294MC provided by ZWO ASTRO) with high sensitivity at a 1-um wavelength, a telescope (SKYMAX127 provided by Sky-Watcher) with a 1,500 mm focal length, and a stealing mirror in the EA chamber.
Here, $\vec{a} = \vec{a}_1 + \vec{a}_2$ is the position of the Pcal center of force, the vector sum of the force vectors for the individual beams.
Since the laser power of the two beams is adjusted to be the same on the ETM, $\vec{a}$ can be considered as the vector sum of the positions of the two beams.
\par
\begin{figure}[ht]
  \centering
  \includegraphics[width=0.6\textwidth]{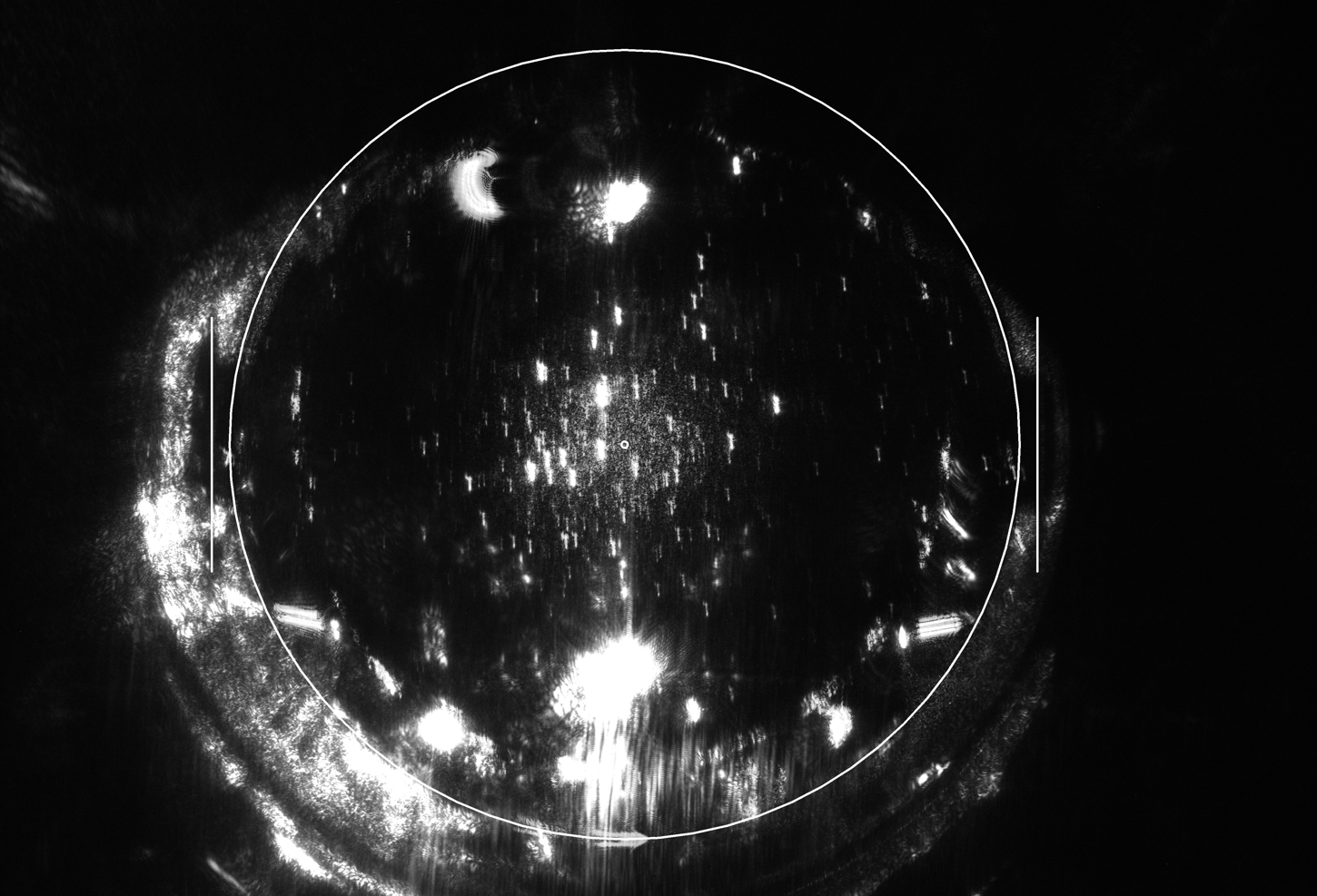}
  \caption{Tcam picture. The mirror face shape and the end test mass mirror (ETM) suspension wires are shown on the picture and were used to estimate the ETM position and the conversion factor (from the number of pixels to mm).}
  \label{fig:Tcam_example}
\end{figure}

 To estimate the beam positions, the ETM position on a picture of the Tcam and a conversion factor $\beta_{\rm Tcam}$ (from the number of pixels in the picture to millimeters) were estimated using the geometrical information of the ETM.
Figure \ref{fig:Tcam_example} shows the Tcam picture data.
Since the shape of the ETM mirror surface and the relative positions of the ETM suspension wires are known, the ETM position on a Tcam picture and $\beta_{\rm Tcam}$ can be estimated.
The error of the estimated ETM position was 10 dots on the image data, corresponding to 0.88 mm (the measured $\beta_{\rm Tcam}$ was 11.3 dots/mm).
\par The beam position of the main interferometer was estimated from a Tcam picture obtained using the normal interferometer operation mode.
In this case, since the laser power in the arm was high ($\sim$ 10 kW), the surface of the ETM viewed by the Tcam was very bright.
Consequently, we used a relatively short exposure for this picture shooting; the camera gain and the exposure time were 100 and 15–25 ms, respectively.
Since the light caught by the Tcam is the scattered light from the ETM surface, the distribution is not uniform, and each scattered point can be bright or dark depending on the properties of the scattered source.
In particular, some very bright points disturbed the image analysis.
Therefore, we used the following procedure to obtain the interferometer beam position:
\begin{enumerate}
    \item Separate into RGB images
    \item Remove saturation: remove pixels with a value of 250 or more.
    \item Apply a Gaussian blur image filter
    \item Apply fitting using a 2D Gaussian function
    \item Calculate the average of the RGB image results.
\end{enumerate}
\par We obtained pictures of the Pcal beams with the interferometer beam not hitting the ETM.
In this case, we used a higher exposure than that used in the interferometer beam position shooting; the camera gain and the exposure time were 470 and 10–100 ms, respectively.
Many bright points, which are contained in the entire image, can be mistaken for Pcal beams; thus, we defined a search area of Pcal beam positions and searched for the brightest point by applying a 2D Gaussian fitting function around the tracked points.
We used the following procedure for the Pcal beam-position estimation:
\begin{enumerate}
    \item Convert an input RGB to a gray image.
    \item Find the brightest point of the Pcal beams in the defined region.
    \item Apply a 2D Gaussian fitting function around the tracked point.
\end{enumerate}
The 2D Gaussian function used in the analysis is the following:

\begin{equation}
    f(x,y) = \frac{A}{2\pi\sigma^2} \exp{\left(-\frac{(x-\mu_x)^2 + (y-\mu_y)^2}{2\sigma^2}\right)}
    \label{eq:2D_Gaussian}
\end{equation}

where $x$ and $y$ are the position coordinates on the figure, and $A, \sigma, \mu_x$, and $\mu_y$ are the normalized amplitude, SD, and estimated beam-position coordinates.

\section{Beam-position estimation (result in O4a)}
\par
\begin{figure}[htbp]
  \centering
  \subfloat{
    \includegraphics[width=0.3\textwidth]{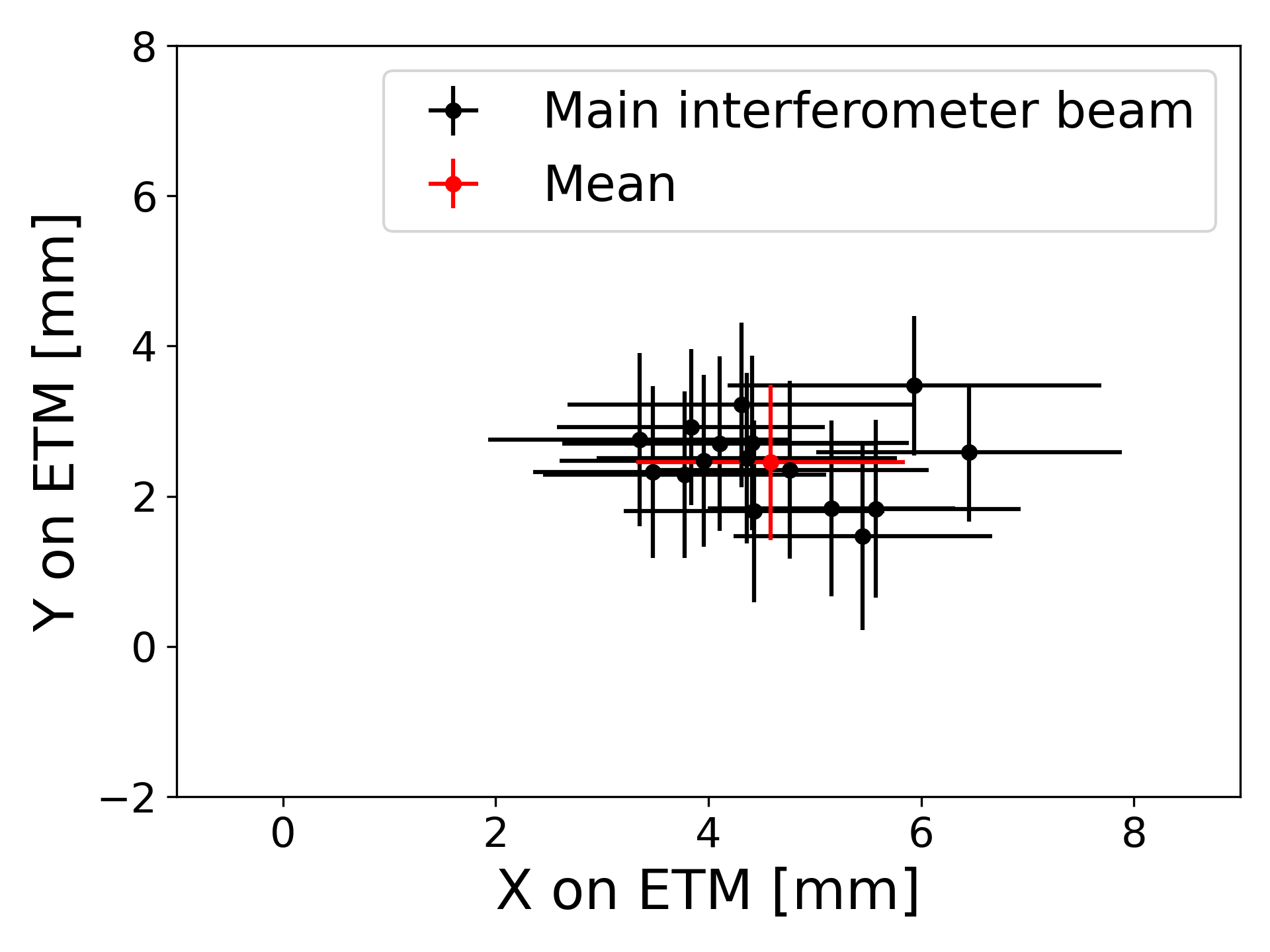}
  }
  \subfloat{
    \includegraphics[width=0.3\textwidth]{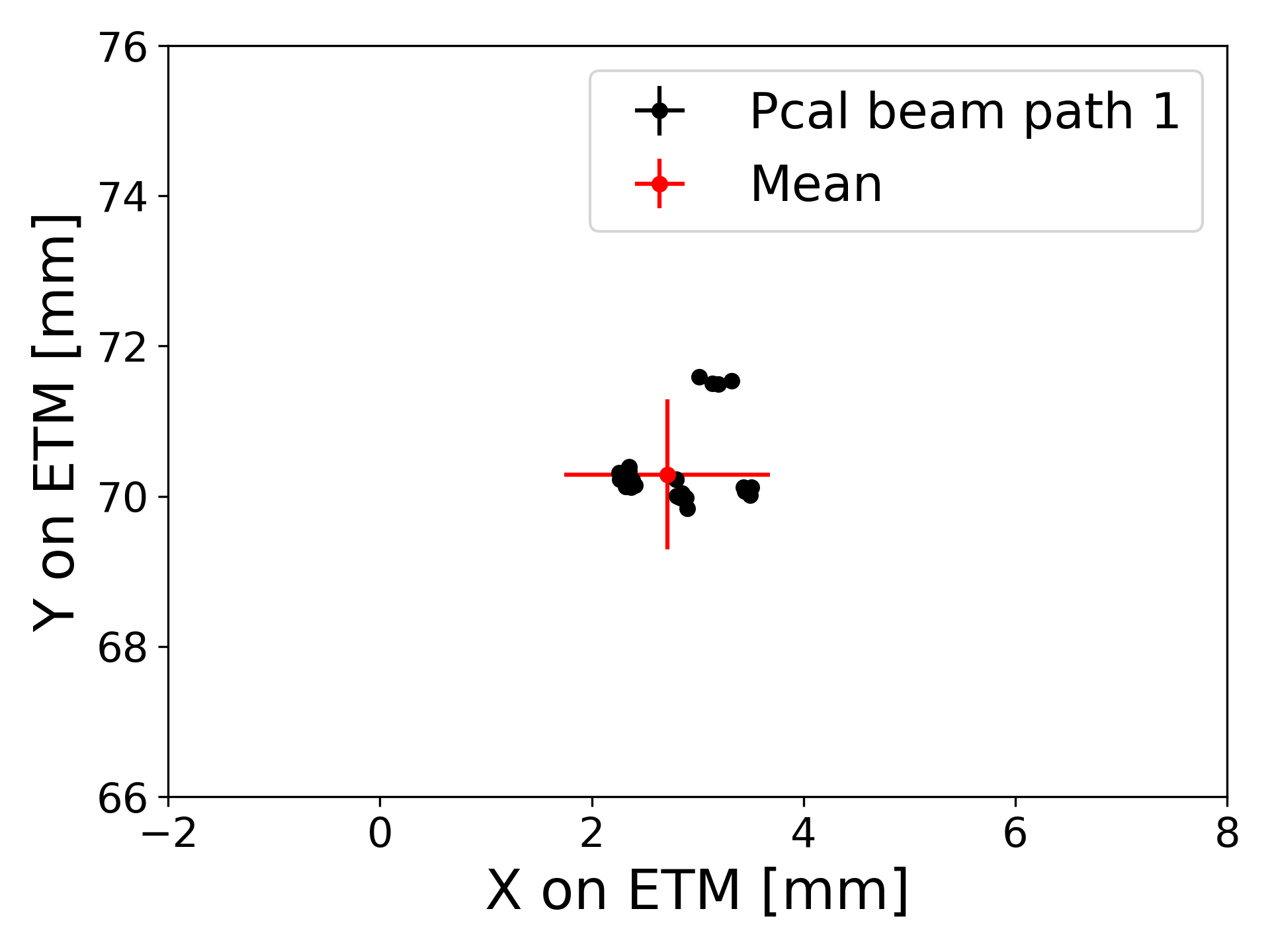}
  }
  \subfloat{
    \includegraphics[width=0.3\textwidth]{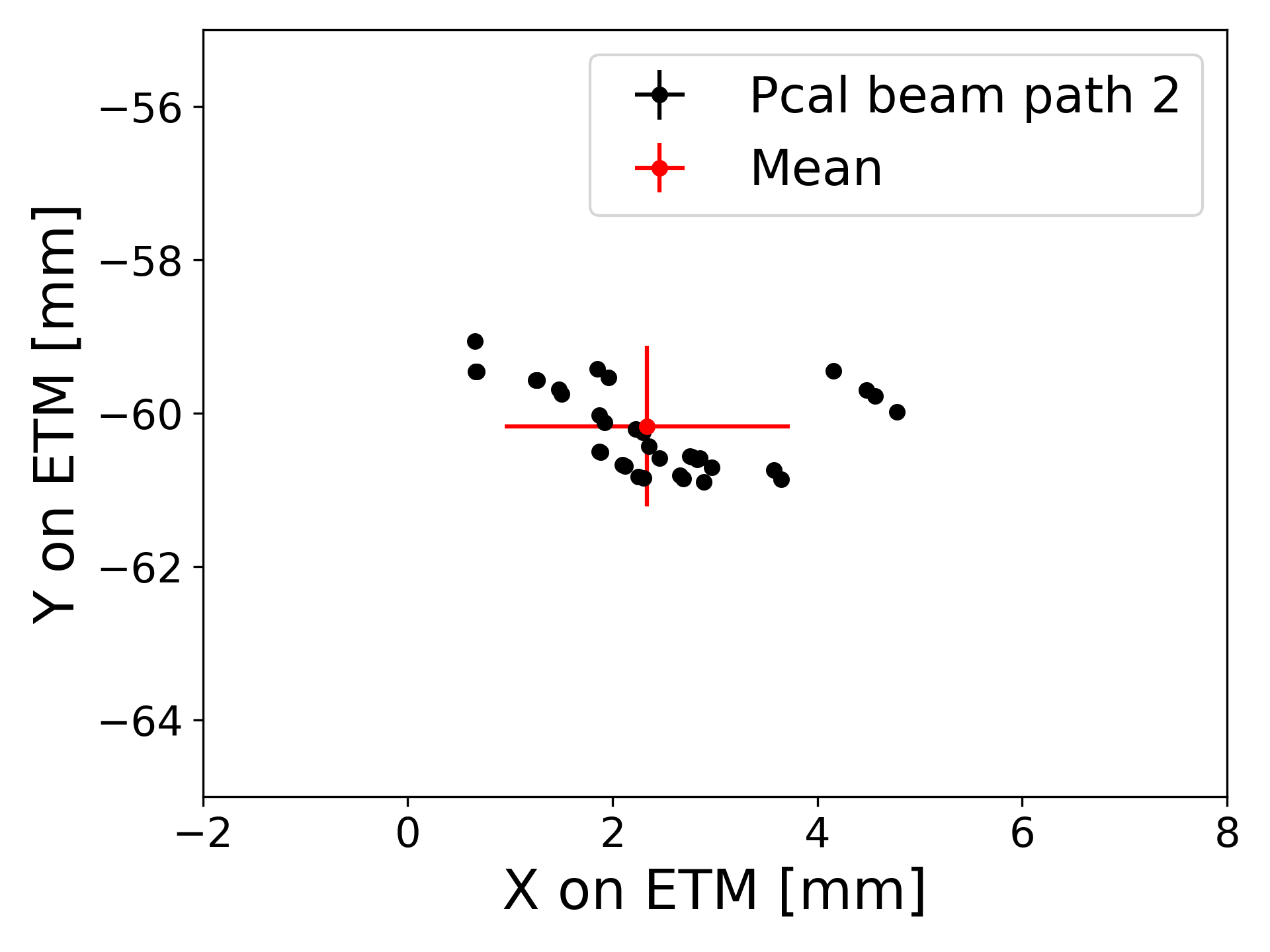}
  }
  \caption{Left: Variation of the main interferometer beam position during O4a. Middle: Variation of the Pcal-X path 1 beam position. Right: Variation of the Pcal-X path 2 beam position. (The coordinate origin is at the center of the ETMX.)}
  \label{fig:Tcam_figures_paper}
\end{figure}

 16 and 37 images were obtained for the main interferometer beam and the Pcal beams, respectively, on maintenance days during O4a.
These images were analyzed to obtain their variation, which corresponds to the variation in the estimated beam position during the O4a period.
The results are shown in Figure \ref{fig:Tcam_figures_paper} and summarized in Table \ref{tab:beam_position}.

\par The values are summarized on the Table \ref{tab:beam_position}.

\begin{table}[htbp]
    \centering
    \begin{tabular}{|c|c|c|c|}
        \hline
        [mm] & Main IR & Pcal-X path 1 & Pcal-X path 2 \\
        \hline
        X & $4.6 \pm 1.3$ & $2.7 \pm 1.0$ & $2.3 \pm 1.4$ \\
        Y & $2.5 \pm 1.0$ & $70.3 \pm 1.0$ & $-60.2 \pm 1.0$ \\
        \hline
    \end{tabular}
    \caption{Estimated beam positions during O4a. The coordinate origin is at the center of the ETMX; the +X and +Y axes indicate the negative y-arm of the interferometer and vertical up directions, respectively.}
    \label{tab:beam_position}
\end{table}

\par These values were used to estimate the ETMX displacement caused by the Pcal and the uncertainty, which is described in the next section.

\section{Total Pcal uncertainty}
\par To estimate the total uncertainty in Pcal, we need to know the uncertainty in each parameter in Equations \ref{eq:Pcal_main}, \ref{eq:Pcal_I}, and \ref{eq:P_TM_estimation}.
\par Specifically, we need to know the mass $M$ of the ETMX, the thickness $t$ and radius $r$ of the ETMX, the Pcal incident angle $\theta$, and the optical efficiencies between the Tx module and ETMX and between ETMX and the Rx module ($e_{\rm T1}$ and $e_{\rm R1}$ and $e_{\rm T2}$ and $e_{\rm R2}$, respectively).
\subsection{Parameters related to the ETMX mirror}
\par The ETMX mass and shape were measured before the installation as follows \cite{doi:10.1142/9789811258251_0237}: 
\begin{eqnarray}
    M &=& 23.01 \pm 0.01 \ [{\rm kg}], \\
    t_{\rm ETMX} &=& 0.14996 \pm 0.000050 \ [{\rm m}], \\
    r_{\rm ETMX} &=& 0.11014 \pm 0.000025 \ [{\rm m}].
\end{eqnarray}
$M$ includes the mass of the mirror itself, the two side ears used for suspension, and the four nail heads of the sapphire fibers.
These values were used to calculate the inertial moment of the ETMX $I$ and the uncertainty.

\subsection{Incident angle of Pcal beams on the ETMX}
\par To estimate the incident angle of the Pcal beams on the ETMX, we measured the distance $d_{s}$ between the interferometer and the Pcal beams and the distance between the EXC chamber, which contains the ETMX, and the EXA chamber, which contains the last mirrors for the Pcal beams $d_{l}$.
\begin{eqnarray}
    d_{s} &=& 0.46 \pm 0.05 \ [{ \rm m}] \\
    d_{l} &=& 34.9 \pm 0.2 \ [{ \rm m}].
\end{eqnarray}
\par Using the above parameters, we estimated the uncertainty in $\cos\theta$ in Equation \ref{eq:Pcal_main} (\ref{appendix:theta}).

\subsection{Total Pcal uncertainty in O4a}
\par In the estimation of the Pcal-X uncertainty in O4a, we used the typical voltage value of the integrating spheres based on the data obtained during O4a. The values used in this calculation are presented in Table \ref{table:IS_values_in_O4a}.
\begin{table}[htbp]
    \centering
    \caption{Typical output voltage values from power-srnsors for Pcal-X during the O4a}
    \begin{tabular}{|c|c|c|c|}
        \hline
         & TxPD 1 & TxPD 2 & RxPD \\
        \hline
        \hline
        Typical output [V] & 1.9499 & 2.6970 & 1.3263 \\
        \hline
    \end{tabular}
    \label{table:IS_values_in_O4a}
\end{table}
\par Using the voltage values given in Table 4, we estimated the laser power at ETMX as follows:
\begin{eqnarray}
    P_{\rm ETMX} &=& 2.009 \pm 0.013 \ (0.67\ \%) \ [{\rm W}]
\end{eqnarray}
\par Other parameter uncertainties were also included in the Pcal-X uncertainty estimation. The following result was obtained:
\begin{equation}
    x_{\rm 1Hz} = (1.490 \pm 0.011)\times 10^{-11} [{\rm m}]
\end{equation}
This value indicates that when the OFS system outputs a 1-Hz sine wave with maximum amplitude and a DC offset, as shown in Table \ref{table:IS_values_in_O4a}, this displacement will be induced in the interferometer.

As a result, the total Pcal-X relative uncertainty in O4a is 0.73\ \%.
\par The final uncertainty contributions of parameters $P, \vec{a}\cdot\vec{b}, M, I$, and $\cos\theta$ were individually evaluated by considering the propagation error associated with each parameter while keeping all the other parameters constant.
The results are presented in Table \ref{table:error_contribution}.

\begin{table}[h!]
    \centering
    \caption{Pcal uncertainty contribution from each parameter (Pcal-X). Relative uncertainty is calculated using the exact values, while the main value and uncertainty are rounded for presentation.}
    \resizebox{\textwidth}{!}{
        \begin{tabular}{|c|c|c|c|c|}
            \hline
            Parameter & Main value & Uncertainty & Relative uncertainty [\%] & Contribution to total [\%] \\
            \hline
            \hline
            $P$ & 2.009 W & 0.013 W & 0.67 & 0.67 \\
            $\vec{a}\cdot\vec{b}$ & $4.8\times10^{-5}\ {\rm m^2}$ & $1.5\times10^{-5}\ {\rm m^2}$ & 31 & 0.30 \\
            $M$ & $23.01\ {\rm kg}$ & $0.01\ {\rm kg}$ & 0.043 & 0.043 \\
            $I$ & $0.112903\ {\rm kg m^2}$ & $0.000056\ {\rm kg m^2}$ & 0.049 & 0.00048 \\
            $\cos\theta$ & $0.999913$ & $0.000019$ & $0.0019$ & $0.0019$ \\
            \hline
            $x_{\rm 1Hz}$ & $1.490\times10^{-11}\ {\rm m}$ & $0.011\times10^{-11}\ {\rm m}$ & 0.73 & - \\
            \hline
        \end{tabular}
    }
    \label{table:error_contribution}
\end{table}

\par As shown in Table \ref{table:error_contribution}, the final relative uncertainty of Pcal-X, which is the main calibration system in KAGRA in O4a was 0.73\ \%.
This is more than three times better than the value of 3\ \% at the time of O3GK \cite{10.1093/ptep/ptab018}.
The laser power and beam-position estimation mostly contribute to the overall Pcal uncertainty.
The relative uncertainty in the incident angle $\theta$ has a small impact on the overall uncertainty.
One of the reasons for this is that the Pcal system is located far from the ETM (34.9 m) to prevent interference with the cryogenic system, such as the cryogenic duct shield in the vacuum tube and the cryogenic coolers around the mirror chamber.
Due to this long baseline, the incidence angle $\theta$ is small, making $\cos \theta$ very close to unity, which further reduces its contribution to the total uncertainty.

\section{Discussion}
\par The 0.73\ \% relative uncertainty in Pcal calculated in this study is one of the uncertainties in reconstructing h(t) to which the instability of the interferometer response and measurements are added to determine the uncertainty of the final reconstructed h(t).
If the uncertainty in h(t) is smaller than the inverse of the signal-to-noise ratio (SNR) of the observed GW signal, it does not significantly impact the estimation of the GW source parameters.
Given the sensitivity of KAGRA in O4 and the SNR of the GW signal obtained by LIGO and Virgo to this day, the 0.73\ \% relative uncertainty is sufficiently low.
\par The largest contribution to the 0.73\ \% estimated relative uncertainty in the overall Pcal is the estimation of the Pcal laser beam power on the ETM; the second largest contribution is the accuracy in identifying the beam position on the ETM.
The uncertainty in the laser power estimation is limited by the fluctuation in the calibration parameters of the power sensors during each calibration measurement day.
As the fluctuation is much higher than the statistical error of each measurement point, we assume that the systematic errors, which depend on how the integrating sphere is placed and measured, are included.
We believe that ensuring the appropriate angle of the laser beam incident on the power sensor and reducing the angular dependence will improve the overall Pcal uncertainty, which is one of our future objectives. Identifying the beam positions has been a challenge to improve the Pcal uncertainty; we demonstrated this using the Tcam.
In this study, we used the data obtained on weekly maintenance days.
In the future, if it is possible to monitor the beam positions (almost) constantly during the observation, we will be able to obtain the time variation of the beam positions and estimate the Pcal uncertainty more accurately.

\section{Conclusion}
\par We presented the configuration and performance of a Pcal system in KAGRA, which is used as the main calibration system.
For a Pcal system that transmits a reference signal to the interferometer by injecting an amplitude-modulated laser beam into an ETM, the laser beam power and the estimation of the laser beam position have the highest impact on the Pcal uncertainty.
We estimated a low overall relative uncertainty of 0.73\ \% in O4a due to the power sensor calibration, which measures the laser power, and the Tcam system, which captures the beam-position photographically.
The successful operation of the Pcal system in KAGRA with its long cryogenic duct shields, which employ low-temperature interferometers, provides useful information for future low-temperature interferometers (ET and CE telescopes).

\section{Acknowledgments}
\par This work was partly supported by JSPS KAKENHI Grant No. 22H00135 (T. Tomaru), No. 17H06133 (N. Kanda), No. 17H01135 (T. Tomaru), and No. 20K22353 (D. Chen). 
It was also supported in part by the Inter-University Research Programs of the Institute for Cosmic Ray Research (ICRR), University of Tokyo. 
Additional support was provided by MEXT, the JSPS Leading-edge Research Infrastructure Program, the JSPS Grant-in-Aid for Specially Promoted Research 26000005 (T. Kajita), and the JSPS Core-to-Core Program A.

\appendix
\section{Average value and uncertainty calculations}\label{appendix:average}
\par We present the method used to calculate the average value and its uncertainty.
\par To compute the average of a vector $\bm{A}$ with $N$ elements, we employed the following weighted average method ($A_i$ $(i=1,\dots, N)$ are the elements of $\bm{A}$, and each element has an associated uncertainty denoted by $\sigma_i$):
\begin{eqnarray}
    \bar{A} &=& \frac{\sum w_i A_i}{\sum w_i}, \\
    \langle \bm{A} \rangle &=& \bar{A} \pm \sqrt{\frac{\sum w_i (A_i - \bar{A})^2}{\frac{N-1}{N} \sum w_i}},
\end{eqnarray}
$w_i$ represents the weight for each data point, defined as follows:
\begin{equation}
    w_i = \frac{1}{\sigma_i^2}
\end{equation}

\par If each component of $\bm{A}$ does not have an associated uncertainty, the arithmetic mean and standard deviation were calculated as follows:
\begin{eqnarray}
    \bar{A} &=& \frac{\sum A_i}{N}, \\
    \langle \bm{A} \rangle &=& \bar{A} \pm \sqrt{\frac{\sum (A_i - \bar{A})^2}{N-1}}
\end{eqnarray}

\section{Calculation of the contribution of the Pcal incident beam angle to the uncertainty}\label{appendix:theta}
\par By measuring the distance $d_l$ between the last mirrors in the EA chamber and the ETMX, and the distance $d_s$ between the last mirrors and the main interferometer laser path, the value and the uncertainty of $\cos \theta$ in Equation (\ref{eq:Pcal_main}) can be calculated as follows:
\begin{eqnarray}
    \cos\theta &=& \frac{d_{l}}{\sqrt{d_{s}^2 + d_{l}^2}}, \\
    \Rightarrow \Delta \cos\theta &=& \sqrt{\left( \frac{d_{s}d_{l}}{\sqrt{d_{s}^2 + d_{l}^2}} \Delta d_{s} \right)^2 + \left( \frac{d_{s}^2}{\sqrt{d_{s}^2 + d_{l}^2}} \Delta d_{l} \right)^2}
\end{eqnarray}


\section*{References}
\bibliography{cite.bib}
\bibliographystyle{plain}
\end{document}